\documentstyle[eqsecnum,aps,amssymb,psfrag,amsmath,graphicx]{revtex}

\newcommand{\U}[1]{\hat{U}_{#1}}
\newcommand{\ket}[1]{\left| #1 \right\rangle}
\newcommand{\bra}[1]{\left\langle #1 \right|}
\newcommand{\tr}[1]{\operatorname*{tr}\left\{ #1 \right\} }

\newcommand{\id}{\hat{1}}
\newcommand{\bib}[4]{\bibitem{#1} {#2}, {#4}}
\newcommand{\vol}[1]{{\bf{#1}}}

\newcommand{\tab}[1]{\framebox(1.0,1.0){\scriptsize #1}}

\newcommand{\tabe}[4]{\setlength{\unitlength}{4mm}\begin{picture}(4,1.5)\put(0.0,0.0){\tab{#1}}\put(1.0,0.0){\tab{#2}}\put(2.0,0.0){\tab{#3}}\put(3.0,0.0){\tab{#4}}\end{picture}}

\newcommand{\tabf}[4]{\setlength{\unitlength}{4mm}\begin{picture}(4,2.5)\put(0.0,1.0){\tab{#1}}\put(1.0,1.0){\tab{#2}}\put(2.0,1.0){\tab{#3}}\put(0.0,0.0){\tab{#4}}\end{picture}}

\newcommand{\tabg}[4]{\setlength{\unitlength}{4mm}\begin{picture}(4,2.5)\put(0.0,1.0){\tab{#1}}\put(1.0,1.0){\tab{#2}}\put(0.0,0.0){\tab{#3}}\put(1.0,0.0){\tab{#4}}\end{picture}}

\begin{document}
\twocolumn
\widetext

\title{Adapted operator representations:\\
Selective versus collective
properties of quantum networks}
\author{Alexander Otte and G\"unter Mahler}
\address{Institut f\"ur Theoretische Physik,\\Universit\"at Stuttgart, Pfaffenwaldring 57,\\70550 Stuttgart, Germany}
\maketitle

\makeatletter
\def\endtable{\global\tableonfalse\global\outertabfalse
{\let\protect\relax\small\vskip2pt\@tablenotes\par}\xdef\@tablenotes{}%
\egroup}%
\def\@sect#1#2#3#4#5#6[#7]#8{\ifnum #2>\c@secnumdepth
\let\@svsec\@empty\else
\refstepcounter{#1}%
\def\@tempa{#8}%
\ifx\@tempa\empty %
\ifappendixon %
\if@mainhead %
\def\@tempa{APPENDIX }\def\@tempb{}%
\else %
\def\@tempa{}\def\@tempb{. }%
\fi
\else %
\def\@tempa{}\def\@tempb{. }%
\fi
\else %
\ifappendixon %
\if@mainhead %
\def\@tempa{APPENDIX }\def\@tempb{: }%
\else %
\def\@tempa{}\def\@tempb{. }%
\fi
\else %
\def\@tempa{}\def\@tempb{. }%
\fi
\fi
\edef\@svsec{\@tempa\csname the#1\endcsname\@tempb}\fi
\@tempskipa #5\relax
\ifdim \@tempskipa>\z@
\begingroup #6\relax
{\hskip #3\relax\@svsec}{\interlinepenalty \@M
\if@mainhead\uppercase{#8}\else#8\fi\par}%
\endgroup
\csname #1mark\endcsname{#7}\addcontentsline
{toc}{#1}{\ifnum #2>\c@secnumdepth \else
\protect\numberline{\csname the#1\endcsname.}\fi
#7}\else
\def\@svsechd{#6\hskip #3\relax %
\@svsec \if@mainhead\uppercase{#8}\else#8\fi
\csname #1mark\endcsname
{#7}\addcontentsline
{toc}{#1}{\ifnum #2>\c@secnumdepth \else
\protect\numberline{\csname the#1\endcsname.}\fi
#7}}\fi
\@xsect{#5}}

\batchmode\font\xyzfont=cmssi8 \errorstopmode
\ifx\xyzfont\nullfont \font\xyzfont=cmr8 \fi
\def\form{\par\vskip-\lastskip\vskip.5pc plus.5pc
\noindent{\small\bf Form:}\vskip1pc plus.25pc}
\global\@specialpagefalse
\def\@oddhead{} \let\@evenhead\@oddhead
\def\centerexample{\par\vskip-\lastskip\vskip1pc \edef\@@@e{\the\c@equation}
  \edef\@@@s{\the\c@section} \edef\@@@ss{\the\c@subsection}
  \edef\@@@sss{\the\c@subsubsection} \edef\@@@t{\the\c@table}
  \edef\@@@f{\the\c@figure} \c@section0 \c@subsection0 \c@subsubsection0
  \c@figure0 \c@table0 \appendixonfalse \noindent\centering
}
\def\endcenterexample{\vskip1pc\global\@ignoretrue
  \global\c@section=\@@@s\relax \global\c@subsection=\@@@ss\relax
  \global\c@subsubsection=\@@@sss\relax \global\c@table\@@@t\relax
  \global\c@figure\@@@f\relax \global\c@equation=\@@@e\relax
  \global\@ignoretrue \noindent
}
\def\example{\par\vskip-\lastskip\vskip1pc \edef\@@@e{\the\c@equation}
  \edef\@@@s{\the\c@section} \edef\@@@ss{\the\c@subsection}
  \edef\@@@sss{\the\c@subsubsection} \edef\@@@t{\the\c@table}
  \edef\@@@f{\the\c@figure} \c@section0 \c@subsection0 \c@subsubsection0
  \c@figure0 \c@table0 \appendixonfalse \noindent
}
\def\endexample{\vskip1pc\global\@ignoretrue
  \global\c@section=\@@@s\relax \global\c@subsection=\@@@ss\relax
  \global\c@subsubsection=\@@@sss\relax \global\c@table\@@@t\relax
  \global\c@figure\@@@f\relax \global\c@equation=\@@@e\relax
  \global\@ignoretrue \noindent
}
\def\smallexample{\par\vskip-\lastskip\vskip1pc\small
  \edef\@@@e{\the\c@equation}
  \edef\@@@s{\the\c@section} \edef\@@@ss{\the\c@subsection}
  \edef\@@@sss{\the\c@subsubsection} \edef\@@@t{\the\c@table}
  \edef\@@@f{\the\c@figure} \c@section0 \c@subsection0 \c@subsubsection0
  \c@figure0 \c@table0 \appendixonfalse \noindent
}
\def\endsmallexample{\vskip1pc\global\@ignoretrue
  \global\c@section=\@@@s\relax \global\c@subsection=\@@@ss\relax
  \global\c@subsubsection=\@@@sss\relax \global\c@table\@@@t\relax
  \global\c@figure\@@@f\relax \global\c@equation=\@@@e\relax
  \global\@ignoretrue \noindent
}
\def\labelenumi{(\theenumi) }
\def\theenumi{\arabic{enumi}}
\def\labelenumii{(\theenumii) }
\def\theenumii{\alph{enumii}}
\def\p@enumii{\theenumi}
\def\labelenumiii{\theenumiii. }
\def\theenumiii{\roman{enumiii}}
\def\p@enumiii{\theenumi(\theenumii)}
\def\labelenumiv{\theenumiv. }
\def\theenumiv{\Alph{enumiv}}
\def\p@enumiv{\p@enumiii\theenumiii}

\def\enumerate{%
\ifnum \@listdepth>5\relax \@toodeep\else\global\advance\@listdepth\@ne \fi
\def\item{
\expandafter\advance\csname c@enum\@roman{\the\@listdepth}\endcsname by1
\par\leavevmode\csname labelenum\@roman{\the\@listdepth}%
          \endcsname\ignorespaces}%
}

\def\endenumerate{\par\global\advance\@listdepth\m@ne\global\@ignoretrue
\noindent}

\def\appendix{\par\global\appendixontrue
\setcounter{section}{0}
\setcounter{subsection}{0}
\setcounter{subsubsection}{0}
\def\thesection{\Alph{section}}
\def\thesubsection{\arabic{subsection}}
\def\thesubsubsection{\alph{subsubsection}}
\def\theequation@prefix{\thesection}
\@addtoreset{equation}{section}
}
\makeatother


\begin{abstract}
Based on local unitary operators acting on a $n$-dimensional
Hilbert-space, we investigate selective and collective operator basis
sets for $N$-particle quantum networks. Selective cluster operators
are used to derive the properties of general cat-states for any $n$
and $N$. Collective operators are conveniently used to account for
permutation symmetry: The respective Hilbert-space dimension is then
only polynomial in $N$ and governed by strong selection rules. These
selection rules can be exploited for the design of decoherence-free
subspaces as well as for the implementation of efficient routes to
entanglement if suspended switching between states of different
symmetry classes could be realized.  
\end{abstract}

\pacs{Valid PACS appear here.
{\tt$\backslash$\string pacs\{\}} should always be input,
even if empty.}

\narrowtext

\vspace{-8mm}
\section{Introduction}
Problem-adapted representations are convenient tools for dealing with
concrete physical models in virtually any branch of physics: Examples
are the choice of coordinates in classical mechanics, of mode
representations in linear wave theory, of state vector- or
matrix-representations and of complete operator basis sets in quantum
mechanics. 
The mode of adaption may refer to the internal symmetry of the system
under consideration and/or its coupling to the outside world (means of
measurement and control). Adapted representations - though in
principle equivalent to any other - are expected to simplify numerical
calculations and to enhance insight. (cf. \cite{27})

Operators in quantum mechanics may represent observables, states or
transformations. In any case it is convenient to think in terms of
"elements", i.e. basic operator sets, out of which any other operator
could be constructed \cite{21}: For a $n$-dimensional Hilbert-space there are
$n^2$ such (orthogonal) basis operators (defining Liouville-space). If
we prefer to think in terms of basic observables or states, these
basis operators should be chosen hermitian; a pertinent example are
the $SU(n)$-generators. If we rather think in terms of basic
"actions", the basis-operators should be unitary (thus defining basic
unitary transformation). The latter approach has become the method of
choice for investigations relating to quantum computation and quantum
information processing \cite{26}: there we are typically concerned with
sequences of unitary transformations. (Only for $n=2$ are unitarity and
hermiticity compatible requirements for a complete basis set.)

Quantum networks (composite systems) may be described in terms of
product-operators, $\hat{Q}$; if each local operator is taken from the
respective unitary basis set, also the product-operators are
unitary. Furthermore, they are completely specified by the type of
"action" to be applied on each subsystem $\mu$, $\mu=1,2,\dots,N$;
these $\hat{Q}$-operators will be termed "selective".
Alternatively, we may introduce operators $\hat{E}$ which specify the
action but not the "address". In this case we are naturally led to
"collective operators", defining a specific action on a given number $\alpha$ of
subsystems.
Complete sets require the inclusion of phases. Collective as
well as selective operator sets are equivalent: in particular, we
can express one type by the other.
A subset of collective operators has permutation symmetry: These are
the only allowed operators for fundamentally indistinguishable
subystems (fermions or bosons).

Typical scenarios realized, e.g. in nanostructures, will exhibit
neither complete selectivity nor complete non-selectivity (permutation
symmetry). However, systems with weakly restricted selectivity should
still more efficiently be described by $\hat{Q}$-Operators, systems
with weakly broken permutation symmetry by collective operators. It
is this latter theme of operational (partial) indistinguishability
which will be of central interest in our present investigation.

Our paper is organized as follows: In section \ref{BasisOperators} we
discuss local basis operators with special emphasis on unitary
operators and introduce the complex coherence vector as a
generalization to the Bloch vector. Section \ref{ClusterOperators}
extends this concept to quantum networks by the use of index-selective
cluster operators. Alternatively, section
\ref{CollectiveOperatorBasis} proposes the use of collective operators
for which all subsystems are treated
on equal footing. Applications of these concepts are
worked out in section \ref{Applications}, starting with highly entangled
states and generalized cat states. Permutation symmetry plays the key
role in the remaining applications showing the fundamental difference
between selective and collective treatment of quantum networks.

\section{Basis operators}
\label{BasisOperators}

\subsection{Transition- and SU($n$)-operators}

The Hilbert-space of dimension $n$ is taken to be spanned by a
complete and orthonormal set of states $\ket{j}$,
$j=0,1,2,\dots,n-1$. They may be considered eigenstates of some
operator $\hat{A}$,
\begin{equation}
\hat{A}=\sum_{j=0}^{n-1}A_j \hat{P}_{jj}
\end{equation}
where $\hat{P}_{ij}=\ket{i}\bra{j};\quad \hat{P}_{ij}^{\dagger} =
\hat{P}_{ji}$ denote $n^2$ transition operators, which are
characterized by two state indices (= quantum numbers = eigenvalues
$A_j$); they are orthonormalized,
$\tr{\hat{P}_{ij}\hat{P}_{kl}^{\dagger}}=\delta_{ik} \delta_{jl}$, (tr = trace operation) and
complete. Any other operator $\hat{B}_s$ can then be written as
\begin{equation}
\hat{B}_s = \sum_{ij} B_{s,ij} \hat{P}_{ij}
\label{trafo}
\end{equation}
where $B_{s,ij}=\tr{\hat{B}_s \hat{P}_{ij}^{\dagger}}$. We require the 
$\hat{B}_s$, $(s=0,1,\ldots,n^2-1)$ to form a complete orthogonal set,
normalized to $n$, 
\begin{equation}
\tr{\hat{B}_s\hat{B}_{s^{\prime}}^{\dagger}}=\sum_{ij}
B_{s,ij}B_{s^{\prime},ij}^{*}= n\, \delta_{ss^{\prime}}\;.
\end{equation}
(Remember that the identity operator $\hat{1}$ is
also normalized to $n$.)
Note, that the transformation eq. (\ref{trafo}) does not change the representation
of any given operator; rather we change the set of operators {\it
within} the $\hat{A}$-representation (i.e. the set of elementary
matrices). A well-known example are the $n^2$ hermitian
SU($n$)-generators $\underline{\hat{\lambda}}=\{\hat{\lambda}_s, s=0,1,\dots,n^2-1\}= \{\hat{1},
\hat{u}_{01}, \hat{u}_{02},\dots,
\hat{u}_{12},\dots,\hat{v}_{01},\dots,\hat{w}_0,\dots,\hat{w}_{n-2}\}$
\begin{eqnarray}
\hat{\lambda}_0 &\equiv& \hat{1} = \sum_{j=0}^{n-1}\hat{P}_{jj}\;,\\
\hat{u}_{ik} &=& \sqrt{\frac{n}{2}} (\hat{P}_{ik}+\hat{P}_{ki})\;,\\
\hat{v}_{ik} &=& \sqrt{\frac{n}{2}} i (\hat{P}_{ik}-\hat{P}_{ki})\;,\\
\hat{w}_l &=& -\sqrt{\frac{n}{(l+1)(l+2)}}(
\hat{P}_{00}+\dots\\
&&+\hat{P}_{ll} - (l+1) \hat{P}_{l+1,l+1})\;,\nonumber
\end{eqnarray}
with $\tr{\hat{\lambda}_s \hat{\lambda}_{s^{\prime}}}=n\,
\delta_{s,s^{\prime}}$ for all $s$, $s^{\prime}$. (The
SU($n$)-generators are often normalized to $2$ rather than $n$ \cite{21}; the
latter choice is more convenient for our purposes.) For
$n=2$ the representation of the $\hat{\lambda}_j={\hat{\lambda}_0,
\hat{u}_{01}, \hat{v}_{01}, \hat{w}_0}$ leads to the well-known Pauli
matrices. For $n>2$ the SU($n$)-algebra tends to loose its
convenience, because the corresponding operators are no longer unitary
and their definition becomes rather unwieldy (cf. \cite{22}).

With the unitary operators to be discussed next, the SU($n$)-operators
share two properties: 
\begin{itemize}
\item[i.]
The set consists of $n^2-1$ traceless
operators $\hat{B}_s$, $s\neq 0$ and the unit-operator $\hat{B}_0=\hat{1}$.
\item[ii.] The operators based on
projection operators $\hat{P}_{jj}$ are  kept separated from the
others.
\end{itemize}

As a result of i., the product-operators of a composite system
(cf. Sect. \ref{ClusterOperators},\ref{CollectiveOperatorBasis} can be
decomposed into a hierachy of $m$-cluster-operators $\hat{Q}$, where
$m$ is the number of indices $s \neq 0$.

As a result of ii., and if $\hat{A}=\hat{H}$, the Hamiltonian of the
system under consideration, the expectation values of the
$\hat{P}_{jj}$ or combinations of those are constants of motion (under
the unitary evolution generated by $\hat{H}$).

\subsection{Unitary operators}
\label{UnitaryOperators}
The operators \cite{16,16z}
\begin{equation}
\hat{U}_{ab}:=\sum_{k=0}^{n-1}\omega_{n}^{bk}\left|  \underline{k+a}\right\rangle
\left\langle k\right|  \;;\qquad\omega_n=e^{\frac{2\pi i}{n}}\;
\label{Udef}
\end{equation}
are defined in a double index notation $a,b\in\lbrack0,n-1]$, where addition modulo $n$ is
denoted by underlining and the constant $\omega_n$ is the $n$-th root
of unity. We note for later reference, that

\begin{equation}
\sum_{k=0}^{n-1}\omega_n^{b k} = n \delta_{0b}\;; \quad
\sum_{b=0}^{n-1}\omega_n^{b (k-k_0)} = n \delta_{k k_0}\;,
\label{omegasum}
\end{equation}

where $\delta_{ij}$ is the Kronecker-delta. The use of two indices leads to an easy interpretation of the action such an
operator has on a state $\ket{k}$,
\begin{equation}
\hat{U}_{ab}\left|  k\right\rangle =\omega_n^{bk}\left|  \underline
{k+a}\right\rangle \;,
\end{equation}
namely $b$ induces a phase shift and $a$ causes a state-shift (for
a concrete example see Appendix A). This set of $n^{2}$ operators defines an
orthonormal and complete operator
basis for the Liouville-space with
\begin{equation}
\operatorname*{tr}\left\{  \hat{U}_{ab}\hat{U}_{cd}^{\dagger}\right\}
=n \delta_{ac}\delta_{bd}\;. \label{orthonormality}
\end{equation}
A operator $\hat{A}$ can be expanded like
\begin{equation}
\hat{A}=\frac{1}{n}\sum_{a,b=0}^{n-1}u_{ab}\hat{U}_{ab}\;;\qquad
u_{ab}:=\operatorname*{tr}\left\{
\hat{U}_{ab}^{\dagger}\hat{A}\right\}  \;. \label{Aexpansion}
\end{equation}
Coming next we will state some of the basic properties these unitary
operators obey. Because $\U{00} = \hat{1}$, the orthonormality
relation (\ref{orthonormality}) implies $\tr{\U{ab}}=0$ for $a,b \neq
0$. This tracelessness will be essential for the definition of
cluster-operators (c.f. Sect. \ref{ClusterOperators}). The adjoint operators are again, barring a phase, members of the set,
\begin{equation}
\hat{U}_{ab}^{\dagger}=\hat{U}_{ab}^{-1}=\omega_n^{ab}\hat{U}_{\underline
{-a},\underline{-b}}\;. \label{dagger}
\end{equation}
A very useful property of these operators is their cyclic symmetry, implying
that any product reduces to just one operator of the set:
\begin{align}
\hat{U}_{ab}\hat{U}_{cd}  &  =\omega_n^{bc}\hat{U}_{\underline{a+c}%
,\underline{b+d}}\\
\hat{U}_{ab}\hat{U}_{cd}^{\dagger}  &  =\omega_n^{cd}\omega_n^{-bc}\hat
{U}_{\underline{a-c},\underline{b-d}}\\
\hat{U}_{cd}^{\dagger}\hat{U}_{ab}  &  =\omega_n^{cd}\omega_n^{-ad}\hat
{U}_{\underline{a-c},\underline{b-d}}\\
\hat{U}_{ab}\hat{U}_{cd}\hat{U}_{ef}  &  =\omega_n^{bc}\omega_n^{(b+d)e}\hat
{U}_{\underline{a+c+e},\underline{b+d+f}} \label{tripleproduct}
\end{align}
It is remarkable to note that the cyclic property allows to use -
instead of the complete set  $\{\U{ab}\}$ - the two operators
$\{\U{0,n-1},\U{n-1,0}\}$ only. All the others can then be
generated as specific product forms, e.g. for $n=3$: 
$\U{00}=(\U{20})^3, \U{10}=(\U{20})^2, \U{22}=\U{20}\cdot\U{02},
\U{01}=(\U{02})^2, \U{12}=(\U{20})^2\cdot\U{02}$, etc. This property is
reminescent of the creation- and destruction operators,
$\hat{a}^{\dagger}, \hat{a}$, conveniently applied to harmonic
oscillator models; it introduces a kind of "non-linearity" if
expectation values of products (cf. eq. (\ref{expvalue})) are
approximated by products of expectation values.

From eq. (\ref{tripleproduct}) and eq. (\ref{dagger}) we conclude that
\begin{eqnarray}
\U{0d}^{\dagger}\cdot\U{ab}\cdot\U{0d} &=& \omega_n^{-da} \U{ab} \\
\U{d0}^{\dagger}\cdot\U{ab}\cdot\U{d0} &=& \omega_n^{bd} \U{ab} 
\end{eqnarray}
i.e. $\U{a0}$ (and their combinations) are invariant under cyclic
permutations of states, the $\U{0a}$ are invariant under cyclic
permutations of phases.

The determinant  of $\U{ab}$ can only take on the values $\pm1$,
\begin{equation}
\operatorname*{det}\hat{U}_{ab}=(-1)^{(a+b)(n-1)}\;,
\end{equation}
and the eigenvalues, consecutively numbered by $k$, all lie on the unit circle
in the complex plane
\begin{equation}
\lambda_{k}^{n}=\omega_n^{bl+ab\frac{n(n-1)}{2}}=(-1)^{ab(n-1)}\omega_n^{bl}\;.
\end{equation}
From the cyclic symmetry the commutation properties are found to be
\begin{align}
\left[  \hat{U}_{ab},\hat{U}_{cd}\right]  _{\pm}  &  =(\omega_n^{bc}\pm
\omega_n^{ad})\hat{U}_{\underline{a+c},\underline{b+d}}\;,\\
\left[  \hat{U}_{ab},\hat{U}_{cd}^{\dagger}\right]  _{\pm}  &  =\omega_n
^{cd}(\omega_n^{-bc}\pm\omega_n^{-ad})\hat{U}_{\underline{a-c},\underline{b-d}}\;,
\end{align}
and the structure constants $f_{ab,cd,ef}$, defined by
\begin{equation}
\left[  \hat{U}_{ab}^{\dagger},\hat{U}_{cd}\right]  _{-}    =\sum
_{e,f=0}^{n-1}f_{ab,cd,ef}\,\hat{U}_{ef}\;,
\end{equation}
follow as
\begin{equation}
f_{ab,cd,ef}    :=\omega_n^{ab}\,(\omega_n^{-bc}-\omega_n^{-ad})\,\delta
_{e,\underline{a+c}}\delta_{f,\underline{b+d}}\;.
\end{equation}
These structure constants are much simpler than those for SU($n$)!
Relations to other basis sets are summarized in Appendix B.

\subsection{Complex coherence vector}
The general state of a quantum-mechanical system is specified by its
density-operator $\hat{\rho}=\sum_{i,j}\rho_{ij} \hat{P}_{ij}$ with
$\rho_{ij}=\tr{\hat{\rho} \hat{P}_{ij}^{\dagger}}$ defining the
respective density matrix. In many areas of physics, especially for
finite-dimensional state spaces, the description
of $\hat{\rho}$ using a coherence vector (as the set of
expectation values of the underlying SU($n$)-operator basis) has shown great power
because of its intuitive, almost "classical" behaviour. For two-level systems
the coherence vector lives in an ordinary 3-dimensional space. In
general, the SU($n$) basis, which is hermitian, leads to a $(n^2-1)$-dimensional vector
consisting of real values, while the unitary operators studied here give
complex elements. 

For the operators $\hat{U}_{ab}$ we will henceforth use a single index
notation $\hat{U}_{i}$ with $i\in\lbrack0,n^{2}-1]$, interpreting $ab$
as the $n$-adic number representing $i$.

Following eq. (\ref{Aexpansion}), we can expand the density operator
$\hat{\rho}$ as
\begin{equation}
\hat{\rho}=\frac{1}{n}\sum_{i=0}^{n^{2}-1}u_{i}\hat{U}_{i}\; \label{expansion}
\end{equation}

and collect the coefficients to define the complex coherence vector%

\begin{align}
\mathbf{u}  &  : =\{u_{i}\quad,\quad0<i<n^{2}-1\}\\
u_{i}  &  :=\left\langle \hat{U}_{i}^{^{\dagger}}\right\rangle
=\operatorname*{tr}\{\hat{U}_{i}^{^{\dagger}}\hat{\rho}\}\quad\in\mathbb{C} \label{expvalue}
\end{align}

where we deliberately excluded $u_{0}$ since it always equals $1$ because of $\hat{U}%
_{0}=\hat{1}$. From the symmetry (cf. eq. (\ref{dagger}))

\begin{equation}
u_{ab}=u_{\underline{-a},\underline{-b}}^{\ast}\; \omega_n^{ab}\;,
\end{equation}

we conclude that there are $n^{2}-1$ independent real parameters forming the
complex coherence vector. The length of $\mathbf{u}$ can be related to
the trace of the density operator $\hat{\rho}$%

\begin{align}
\left|  \mathbf{u}\right|  ^{2}=n\operatorname*{tr}\{{\hat{\rho}}^{2}\}-1\;,
\end{align}

thus giving a simple criterion to distinguish between pure ($\operatorname*{tr}%
\{{\hat{\rho}}^{2}\}=1$) and mixed ($\operatorname*{tr}\{{\hat{\rho}}^{2}%
\}<1$) states:%

\begin{align}
\left|  \mathbf{u}\right|  ^{2}=n-1  &  \qquad\mbox{pure state}\\
0\leq\left|  \mathbf{u}\right|  ^{2}<n-1  &  \qquad\mbox{mixed state}%
\label{BlochLength}
\end{align}

The convenience of the well known Bloch vector formalism comes from the simple
motion performed by the vector under unitary transformations: Since the length
$\left|  \mathbf{u}\right|  $ is preserved, any unitary time evolution
operator $\hat{U}(t)$ just causes a rotation of the coherence vector. If the
density matrix evolves like $\hat{\rho}(t)=\hat{U}(t)\hat{\rho}(0)\hat
{U}(t)^{\dagger}$, one can easily show that the motion of $\mathbf{u}$ is%

\begin{equation}
\mathbf{u}(t)=\text{T}(t)\;\mathbf{u}(0)\;,
\end{equation}

with rotation matrix $\text{T}(t)$ \cite{28},

\begin{equation}
T_{ij}(t):=\frac{1}{n}\operatorname*{tr}\{\hat{U}_{j}^{\dagger}\hat
{U}(t)^{\dagger}{\hat{U}_{i}^{\dagger}}\hat{U}(t)\}\;.
\end{equation}
Using the Liouville equation one can replace the time evolution operator by
the Hamiltonian $\hat{H}(t)$ leading to a differential equation
similar to the Bloch equations,

\begin{equation}
\dot{{\mathbf{u}}}={\mathbf{\Omega}}\;{\mathbf{u}}\;;\qquad\Omega_{ij}:=-\frac
{1}{ni\hbar}\operatorname*{tr}\{\hat{H}(t)\,\left[  \hat{U}_{i}^{^{\dagger}%
},\hat{U}_{j}\right]  _{-}\}\;.
\end{equation}

As expected, ${\mathbf{\Omega}}$ with its properties $\Omega_{ij} =-\,\Omega
_{ji}^{\ast}$ and $\operatorname*{tr}\{\Omega_{ij}\}=0$ describes a
rotation in complex vector-space.

\section{Cluster-operator basis}
\label{ClusterOperators}

\subsection{Definition and properties}
\label{defprop}

Up to now we have restricted ourselves to a single system, respective one node of a quantum
network. To describe a network of $N$ subsystems, $N>1$, looking at each node
separately is not enough since correlations between nodes emerge -- so the
concept of clusters has to be introduced. A cluster operator acting on $m$
particles is build out of $m$ one particle operators $\hat{U}_{i}^{(\mu)}%
$, $i\neq 0$, where greek indices label different nodes of the network. All cluster operators

\begin{eqnarray}
\hat{U}_{i}^{(\mu)} &:=& \hat{1}\otimes\ldots
\otimes\hat{1}\otimes\underbrace{\hat{U}_i}_{\mbox{node $\mu$}}
\otimes\hat{1}\otimes\ldots\otimes\hat{1}\;,\label{U1}\\
\hat{U}_{ij}^{(\mu\nu)} &:=& \hat{1}\otimes\ldots
\underbrace{\hat{U}_i}_{\mbox{node $\mu$}}
\ldots\otimes\hat{1}\otimes\ldots
\underbrace{\hat{U}_{j}}_{\mbox{node $\nu$}}
\ldots\otimes\hat{1}\label{U2}\\\
&&\mbox{\rm etc.}\nonumber
\end{eqnarray}

are unitary and orthonormal, and together span the complete Liouville-space of the quantum network. Acting on only $m$ nodes out of $N$ means leaving the others
unaffected by choosing $\hat{U}_{0}=\hat{1}$ for them (Thus $m$ is the number of indices $i, j, \dots$ unequal zero).
The general decomposition of $\hat{\rho}$ (like of any other operator
$\hat{A}$) for a $N$ particle quantum network with node $\nu$
being a $n_{\nu}$-level system reads%

\begin{eqnarray}
\hat{\rho}   =\frac{1}{\prod\limits_{\mu=1}^{N}n_{\mu}}\left(  \hat{1}+\sum\limits_{\mu
=1}^{N}\sum\limits_{i=1}^{n_{\mu}^{2}-1}u_{i}^{(\mu)}\hat{U}_{i}^{(\mu)}+\sum\limits
_{\mu<\nu}\sum\limits_{i,j}u_{ij}^{(\mu\nu)}\hat{U}_{ij}^{(\mu\nu)}\right.\nonumber \\
+\hspace{-3mm}\sum\limits_{\mu<\nu<\sigma}\sum\limits_{i,j,k}u_{ijk}^{(\mu\nu\sigma)}\hat
{U}_{ijk}^{(\mu\nu\sigma)}+\ldots+\hspace{-3mm}\sum\limits_{\underbrace{\scriptstyle
i,j,k,\ldots,l}_{N\text{ indices}}}u_{ijk\ldots l}^{(12\ldots
N)}\hat{U}_{ijk\ldots l}^{(12\ldots N)}\Biggr)
\;,\nonumber
\end{eqnarray}
\begin{equation}
\end{equation}


\narrowtext

with the index-selective expectation values

\begin{align}
u_{ij\ldots k}^{(\mu\nu\ldots\sigma)}:=\operatorname*{tr}\left\{  \hat{\rho
}\cdot\hat{U}_{ij\ldots k}^{(\mu\nu\ldots\sigma)^{\dagger}}\right\}
=\left\langle \hat{U}_{ij\ldots k}^{(\mu\nu\ldots\sigma)^{\dagger}%
}\right\rangle \;.
\label{uijk}
\end{align}

The local coherence vectors discussed in Sect. \ref{UnitaryOperators} just turn out to be the correlation tensors of first
order, $m=1$. If the network is in a product state, all correlation
tensors of higher order factor into
a product of local coherence vectors,

\begin{equation}
u_{ij\ldots k}^{(\mu\nu\ldots\sigma)}%
=u_{i}^{(\mu)}u_{j}^{(\nu)}\ldots u_{k}^{(\sigma)}\;
\label{factoring}
\end{equation} 
i.e. these states are completely determined by local properties.
In case of a single subsystem, the length of its coherence vector $\sum_i {|u_i|}^2$ can be
identified with a scalar constant of motion under unitary evaluation. For higher order correlation
tensors, the concept of cluster sums leads to new invariants. For any $m$
particle cluster, the cluster sum is defined as the weight of the
tensor (cf. \cite{21})

\begin{equation}
Y_{m}^{(\mu\nu\ldots\sigma)}:=\sum_{\underbrace{\scriptstyle i,j,\ldots
,k}_{m\text{ indices}\neq 0}}\left|  u_{ij\ldots k}^{(\mu\nu\ldots\sigma)}\right|
^{2}\;,
\end{equation}

with special cases $Y_{0}=1$ and $Y^{(\mu)}_{1}=\left|  \mathbf{u}^{(\mu
)}\right|  ^{2}$. For a network with $N$ nodes, these cluster sums give $2^{N}$
scalar invariants under (products of) local unitary
transformations. All the cluster sums together are
subject to the sum rule

\begin{eqnarray}
\operatorname*{tr}\left\{  \hat{\rho}^{2}\right\}  \prod_{\mu=1}^{N}n_{\mu
}&=&Y_{0}+\sum_{\mu=1}^{N}Y_{1}^{(\mu)}+\sum_{\mu<\nu}Y_{2}^{(\mu\nu)}+\label{sumrule}\\
&&\sum
_{\mu<\nu<\sigma}Y_{3}^{(\mu\nu\sigma)}+\ldots+Y_{N}^{(123\ldots
N)}\;. \nonumber
\end{eqnarray}

For a pure state, $\tr{\hat{\rho}^2}=1$ and for $n_{\mu}=n$ the left hand
side is $n^N$. According to eq. (\ref{factoring}), any $m$-cluster sum
of a product state 
factorizes into its $1$-cluster-sum components. A given cluster of
size $m$ can then be tested as being
in a non-product-state, if there is some partition into smaller
clusters, the cluster-sum product of which is smaller than $Y_m$,
e.g. $Y_1^{(1)} Y_1^{(2)} < Y_2^{(12)}$. The cluster sums can be
related to the "purity" of a $m$ particle
cluster by interpreting the cluster as a $n^{m}$-level system with
coherence vector $\mathbf{u}_m$. (For
simplicity we assume $n_{\mu}=n$.) A normalized "purity
factor" can now be defined from the respective coherence vector length
as $(m\le N)$

\begin{align}
p_{m} &  :=\frac{\left|  \mathbf{u}_{m}\right|  ^{2}}{\left|  \mathbf{u}%
_{m}\right|  _{\text{max}}^{2}}=\frac{n^{m}{\operatorname*{tr}}\left\{
\hat{\rho}_{\text{Cluster}}^{2}\right\}  -1}{n^{m}-1}\label{PurityFactor}\\
&  =\frac{1}{n^{m}-1}\left(  \sum_{\mu=1}^{m}Y_{1}^{(\mu)}+\sum_{\mu<\nu}%
Y_{2}^{(\mu\nu)}+\ldots+Y_{m}^{(12\ldots m)}\right)  \;,
\nonumber
\end{align}

characterizing the purity of a $m$-node cluster on the scale (cf. eq. (\ref{BlochLength}))

\begin{align}
0  &  \leq p_{m}\leq1\\
p_{m}  &  =0\quad\Longleftrightarrow\quad\text{maximal mixed }m\text{-cluster}%
\nonumber\\
p_{m}  &  =1\quad\Longleftrightarrow\quad\text{pure }m\text{-cluster.}\nonumber
\end{align}
Alternatively, the purity could be characterized by the respective
cluster-entropy \cite{29}; $p_m$ has the advantage of being a simple algebraic
function of the expectation values (matrix-elements of the reduced
density matrix).

\section{Collective operator-basis for two-level subsystems}
\label{CollectiveOperatorBasis}

\subsection{Definition and properties}
\label{CollectiveOperators}

For now we will stick to quantum networks
build out of $N$ two level systems, knowing that a generalization to $n$-level
systems is straight forward. First one needs to specify which single
particle operator basis is used, e.g. $\hat{u}_{01}=\hat{\sigma}_{x}$,
$\hat{v}_{01}=\hat{\sigma}_{y}$, $\hat{w}_0=\hat{\sigma}_{z}$ (which
is hermitian and unitary)
or $\hat{\sigma}_{\pm}=\hat{\sigma}_x\pm i \hat{\sigma}_y$, $\hat{\sigma}_{z}$ or any other
complete basis. We then relabel the cluster operators $\hat{U}_{ij\ldots
k}^{(12\ldots N)}$ as $\hat{C}_{\alpha\beta\gamma,p}$, meaning an operator of
dimension $2^{2N}$, where $\alpha$,$\beta$ and $\gamma$ specify the
multiplicity of $\hat{\sigma}_x$, $\hat{\sigma}_y$, and
$\hat{\sigma}_z$, respectively ($\alpha+\beta+\gamma\le N$). Index $p$
specifies a permutation of
these among the $N$ subsystems. The number of
such permutations and hence the index range for $p\in\left[  0,\Omega-1\right]  $ is%

\begin{equation}
\Omega(\alpha,\beta,\gamma)= \frac{N!}{\alpha! \beta! \gamma! (N-\alpha-\beta-\gamma)!}\;.
\end{equation}

These operators $\hat{C}_{\alpha\beta\gamma,p}$, again,  span the whole Liouville-space by defining all subsystem
specific properties. We now go on to define
collective operators $\hat{E}$ by

\begin{equation}
\hat{E}_{\alpha\beta\gamma,b}=\sum_{p=0}^{\Omega-1}\omega_{\Omega}^{pb}\hat{C}_{\alpha\beta\gamma,p}%
\;;\qquad\omega_{\Omega}=e^{\frac{2\pi i}{\Omega}}\;.\label{defineE}
\end{equation}

To ensure that all subsystems are treated on  equal footing, the sum
extends over all permutations $p$, weighted only with pure phase fators, where
$b\in\left[  0,\Omega-1\right]  $ labels the phase shift between
"neighbouring" $p$. (Here the numbering of permutations is a
matter of choice and the phase has no physical meaning.) The set of
collective operators is orthonormal%

\begin{equation}
\frac{1}{\Omega\,2^{N}}\operatorname*{tr}\left\{  \hat{E}_{\alpha\beta\gamma,b}%
\hat{E}_{\alpha^{\prime}\beta^{\prime}\gamma^{\prime},b^{\prime}}^{\dagger}\right\}
=\delta_{\alpha\alpha^{\prime}}\delta_{\beta\beta^{\prime}}\delta_{\gamma\gamma^{\prime}}\delta
_{bb^{\prime}}%
\end{equation}

and complete, so that the density operator $\hat{\rho}$ of the network
can be decomposed as 

\begin{eqnarray}
\hat{\rho}&=&\frac{1}{2^{N}}\sum_{\{\alpha\beta\gamma\}}\sum_{b}E_{\alpha\beta\gamma,b}\hat{E}_{\alpha\beta\gamma,b}\;,\\
E_{\alpha\beta\gamma,b}&=&\frac{1}{\Omega}\operatorname*{tr}\left\{  \hat{\rho}\cdot
\hat{E}_{\alpha\beta\gamma,b}^{\dagger}\right\}  \;.
\end{eqnarray}

The expectation values $E_{\alpha\beta\gamma,b}$ are collective in the
sense that they do {\it not} refer to specific subsystem-indices
(as opposed to the $\hat{U}$-operators, cf. eq. (\ref{U1},\ref{U2})).
The inverse transformation is given by
\begin{equation}
\sum_{b=0}^{\Omega-1}\omega_{\Omega}^{-b
p_0}\hat{E}_{\alpha\beta\gamma,b}=\sum_{p=0}^{\Omega-1}\hat{C}_{\alpha\beta\gamma,b}\sum_{b=0}^{\Omega-1}\omega_{\Omega}^{(p-p_0)b}=\Omega
\hat{C}_{\alpha\beta\gamma,p_0}\;,
\end{equation}
where we have made use of eq. (\ref{omegasum}).
For example, the selective operator $\hat{C}_{100,0}\equiv
\hat{\lambda}_1^{(1)}\otimes\hat{1}\otimes\ldots$ could be written as
$\frac{1}{\Omega}\sum_b^{\Omega-1}\hat{E}_{100,b}$. Note that the
operators $\hat{E}$ are, in general, no longer unitary; for $b=0$ they
are hermitian if the operators $\hat{C}_{\alpha\beta\gamma,p}$ are.

\subsection{Alternative sets}

The operators $\hat{E}$ still distinguish between all three basic
operators. Further reductions are possible: one such variant is (based on $\hat{\sigma}_+$, $\hat{\sigma}_-$, $\hat{\sigma}_z$)

\begin{equation}
\hat{F}_{z,\gamma,b}=\sum_{p=0}^{\Omega(z,\gamma)}\sum_{\alpha,\beta}^{\alpha-\beta=z}\omega_{\Omega(z,\gamma)}^{b
p} \hat{C}_{\alpha\beta\gamma,p}
\label{Fdef}
\end{equation}
which has been considered for nuclear spin-networks. In NMR the
$\hat{F}_z$-terms in a Hamiltonian are said to induce a total of $z$
quantum-"flip-flops" \cite{19}; an $\hat{F}_z$-term entering the
density-operator $\hat{\rho}$ describes the respective coherence order
$|z|$.


Another possibility is

\begin{equation}
\hat{G}_{m,b}=\sum_{p=0}^{\Omega(m)}\sum_{\alpha,\beta,\gamma}^{\alpha+\beta+\gamma=m}\omega_{\Omega(m)}^{b
p} \hat{C}_{\alpha\beta\gamma,p}
\label{Gdef}
\end{equation}
where $p$ are permutations of the $m$ operators (of any type) on
the $N$ subsystems, $m=0,1,\dots,N$. Also this set is still
complete. 

\section{Applications}
\label{Applications}
\subsection{Commuting operator sets and generalized cat basis}
\subsubsection{Commuting sets of cluster operators}

Cluster sums and purity factors can be used to classify states by their non-local
properties. Highly entangled states tend to share correlations among all nodes
rather than between only a few. Bell states \cite{31} ($N=2$) are perfect in this sense, because
they are in a totally mixed state ($Y_{1}=p_{1}=0$) locally and pure otherwise
($Y_{2}=3,p_{2}=1$). Generalizing this point, one may define "highly entangled
states" as states that have a cluster sum distribution with a strong focus on
multi-particle correlations. This definition does not give us a quantitative
measure for multi-particle entanglement, but it can assist us in the search
for new states with correlations of high order. Using the unitary operators
defined above, we can give a constructive way of finding such
states. We restrict ourselves to $n_{\mu}=n$. \newline

\textit{Lemma 1}\newline 
Let $\ket{\psi}$ be an eigenstate of $n^N$ completely commuting
unitary operators out of which there are $q_N$
$N$-cluster-opererators. Then its highest order cluster sum is given
by $Y_N=q_N$.\newline

Note that the modulus of any eigenvalue of any cluster-operator
$U_{ij\dots}^{(\mu\nu\dots)}$ is exactly $1$; the $q_N$ commuting
$N$-cluster-operators thus imply $Y_N=q_N$ provided all the non-commuting
$N$-cluster-operators contribute zero. This must be the case, however,
if there
is a total of $n^N$ commuting operators exploiting the sum rule eq. (\ref{sumrule}).

A set of completely commuting operators is a set in which each operator
commutes with all others. This lemma reduces the problem of finding a state
with maximum cluster sum $Y_{N}$ to the problem of finding a maximum set of commuting
$N$-cluster-operators. Before we can make statements about cluster
operators, the commuting relations of single particle unitary operators
$\hat{U}_{i}$ need to be examined. Since $\left[  \hat{U}_{ab}^{\dagger
},\hat{U}_{cd}\right]  _{-}=0$ iff $\underline{ad-bc}=0$, one can show:\newline

\textit{Lemma 2. (based on theory of congruency classes)}\newline 
Each operator $\hat{U}_{ab}$ commutes
with $n\,gcd(a,b,n)$ other basis operators and to each operator $\hat{U}_{ab}$
there exists a trivial set of $n$ completely commuting basis operators. (gcd =
greatest common divisor) \newline

One of the $n$ commuting operators in the completely commuting set is $\hat
{U}_{0}=\hat{1}$, so the maximal number of commuting one particle operators is
$n-1$.

For a network of $N$ particles with $n$-levels each, the properties of unitary
cluster operators lead to the commutator relation

\begin{align}
\left[  \hat{U}_{a_{1}b_{1}}\otimes\hat{U}_{a_{2}b_{2}}\otimes
...,\hat{U}_{c_{1}d_{1}}\otimes\hat{U}_{c_{2}d_{2}}\otimes...\right]  _{-} &
=0\Longleftrightarrow\nonumber\\
\underline{a_{1}d_{1}-b_{1}c_{1}+a_{2}d_{2}-b_{2}c_{2}+...} &  =0\;,
\end{align}

and from this we find the following statements:

\begin{itemize}
\item [ A]There exists a set of ${(n-1)}^{N}$ completely commuting cluster
operators of size $N$. This set can be constructed from all combinations of
$n-1$ completely commuting one particle cluster operators.

\item[ B] There is a set of
\begin{align}
(n^{2}-1)^{\frac{N}{2}}\text{,\qquad}N\text{ even}\\
(n^{2}-1)^{\frac{N-1}{2}\,}\,(n-1)\text{,\qquad}N\text{ odd}
\label{numelements}
\end{align}

completely commuting $N$ particle cluster operators of the form
\begin{equation}
\hat{U}_{a_{1}b_{1}}\otimes\hat{U}_{b_{1}a_{1}}\otimes\hat{U}_{a_{2}b_{2}%
}\otimes\hat{U}_{b_{2}a_{2}}...\;.
\end{equation}

\item[ C] There can be larger sets of completely commuting operators
than given by eq. (\ref{numelements}). However, no general
constructive method to find them is known to us so far. For $n>2$ we
restrict ourselves to $N\le 3$.

\item[ D] The maximum order of any such set is constrained by $n^{N}-1$. This
maximum value is typically not reached for $N>2$.
\end{itemize}

The results of A, B, C and D are summarized in Table \ref{csum}.

\begin{table}
\begin{tabular}[c]{ll}
\begin{tabular*}{4cm}[c]{@{\extracolsep{\fill}}|l||l|l|l|l|l|}\hline
\multicolumn{6}{|l|}{$n=2$}\\\hline\hline
$N$ & A & B & C & D & Cat\\\hline
1 & 1 & 1 & 1 & 1 & 1\\
2 & 1 & 3 & 3 & 3 & 3\\
3 & 1 & 3 & 4 & 7 & 4\\
4 & 1 & 9 & 9 & 15 & 9\\
5 & 1 & 9 & 16 & 31 & 16\\
6 & 1 & 27 & 33 & 63 & 33\\\hline
\end{tabular*}&
\begin{tabular*}{4cm}
[c]{@{\extracolsep{\fill}}|l||l|l|l|l|l|}\hline
\multicolumn{6}{|l|}{$n=3$}\\\hline\hline
$N$ & A & B & C & D & Cat\\\hline
1 & 2 & 2 & 2 & 2 & 2\\
2 & 4 & 8 & 8 & 8 & 8\\
3 & 8 & 16 & 20 & 26 & 20\\
4 & 16 & 64 & ? & 80 & 60\\
5 & 32 & 128 & ? & 242 & 172\\
6 & 64 & 512 & ? & 728 & 508\\\hline
\end{tabular*}
\\
&\\
\multicolumn{2}{c}{
\begin{tabular*}{4cm}
[c]{@{\extracolsep{\fill}}|l||l|l|l|l|l|}\hline
\multicolumn{6}{|l|}{$n=4$}\\\hline\hline
$N$ & A & B & C & D & Cat\\\hline
1 & 3 & 3 & 3 & 3 & 3\\
2 & 9 & 15 & 15 & 15 & 15\\
3 & 27 & 45 & 54 & 63 & 54\\
4 & 81 & 175 & ? & 255 & 213\\
5 & 243 & 525 & ? & 1023 & 828\\
&  &  &  &  & \\\hline
\end{tabular*}}
\end{tabular}
\caption{Cluster sums $Y_{N}$ for eigenstates corresponding to sets of
completely commuting $N$ particle cluster operators calculated by methods A, B
and C. Column D gives the upper limit and colum Cat states the cluster sum
$Y_{N}$ for generalized cat states as introduced in Sect. \ref{generalizedCats}.}%
\label{csum}
\end{table}

\subsubsection{Generalized cat states}
\label{generalizedCats}

As a first example let us look at the well known case of two spin $1/2$
particles. The maximum order for completely commuting sets of
$2$-cluster-operators is 3 and there are
6 such sets:%

\begin{align}
\left\{  \hat{U}_{01}\otimes\hat{U}_{01},\;\hat{U}_{10} \otimes\hat{U}%
_{10},\;\hat{U}_{11}\otimes\hat{U}_{11}\right\}\;,\\
\left\{  \hat{U}_{01}\otimes\hat{U}_{01},\;\hat{U}_{10} \otimes\hat{U}_{11},\;\hat{U}_{11}\otimes\hat{U}_{10}\right\}\;,\nonumber\\
\left\{  \hat{U}_{01}\otimes\hat{U}_{10},\;\hat{U}_{10} \otimes\hat{U}%
_{01},\;\hat{U}_{11}\otimes\hat{U}_{11}\right\}\;,\nonumber\\
\left\{  \hat{U}%
_{01}\otimes\hat{U}_{10},\;\hat{U}_{10} \otimes\hat{U}_{11},\;\hat{U}%
_{11}\otimes\hat{U}_{01}\right\}\;,\nonumber \\
\left\{  \hat{U}_{01}\otimes\hat{U}_{11},\;\hat{U}_{10} \otimes\hat{U}%
_{01},\;\hat{U}_{11}\otimes\hat{U}_{10}\right\}\;,\nonumber\\
\left\{  \hat{U}%
_{01}\otimes\hat{U}_{11},\;\hat{U}_{10} \otimes\hat{U}_{10},\;\hat{U}%
_{11}\otimes\hat{U}_{01}\right\}\;.\nonumber
\end{align}

The eigenstates to the first set form the Bell basis $\left|  00\right\rangle
\pm\left|  11\right\rangle $, $\left|  01\right\rangle \pm\left|
10\right\rangle $, the second set has $i\,\left|  00\right\rangle
\pm\left|  11\right\rangle $, $i\,\left|  01\right\rangle \pm\left|
10\right\rangle $ as eigenstates and the third gives $\,\left|
00\right\rangle +\left|  01\right\rangle +\left|  10\right\rangle -\left|
11\right\rangle $,\ $\,\left|  00\right\rangle +\left|  01\right\rangle
-\left|  10\right\rangle +\left|  11\right\rangle $,\ $\,\left|
00\right\rangle -\left|  01\right\rangle +\left|  10\right\rangle +\left|
11\right\rangle $,\ $\,-\left|  00\right\rangle +\left|  01\right\rangle
+\left|  10\right\rangle +\left|  11\right\rangle $. The remaining sets have
eigenstates similar to the ones given by the third set, except for a phase. All these
states are totally mixed (purity factor $p_{1}=0$)
locally, and pure in total ($p_{2}=1$). From the table for the cluster sums,
one can see that in the case of spin $1/2$ particles ($n=2$) there are no
states with more entanglement than the generalized cat states (see next
chapter). For $n=3$-level systems this is also the case for up to 3 particles, but
in the case of 4 particles there exists a state with a higher cluster sum
$Y_{4}=64$. It turns out that this state is the dyadic product of two Bell
states (each with $Y_{2}=8$). This indicates, that a cat state may not
necessarily have the largest $Y_N$-value
(cf. eq. (\ref{maxYN})).

We generalize
the Bell basis to systems with an arbitrary number of $N$ particles where each
particle is a $n$-level subsystem. Our starting point is the state $\left|
Cat\right\rangle _{0}:=\frac{1}{\sqrt{n}}\sum_{i=0}^{n-1}\underset
{k=1}{\overset{N}{\otimes}}\left|  i\right\rangle $ which has the nice
property, that any subsystem of $m$ particles has an entanglement of
$S(\hat{\rho}_{m})=\log_{2}n$, when measured in terms of the local von
Neumann entropy. ($\hat{\rho}_m $ is the reduced density-operator of
the respective $m$-cluster.) So
any cluster, regardless of its size $m$, $(m<N)$ has the
same entropy and furthermore this entropy is the maximum amount of
entanglement a $n$-level system can have. In this sense this state can be
considered as a maximal entangled state. Since entanglement properties are
invariant under local unitary transformations \cite{28}, one can generate a complete
basis set from $\left|  Cat\right\rangle _{0}$ by applying the discrete set of
unitary basis operators $\hat{U}_{i}$. The result,

\begin{equation}
\left|  Cat\right\rangle _{\mathbf{c}} := \frac{1}{\sqrt{n}}\,{\sum
_{j=0}^{n-1}} \omega_n^{j\,c_{1}}\,\left|  j\right\rangle \underset
{k=2}{\overset{N}{\otimes}}\left|  \underline{j+c_{k}}\right\rangle\;,\label{catdef}
\end{equation}

is an explicit definition of the 
cat state basis. The index
${\mathbf{c}}=\{c_{i}\},\;c_{i}\in\lbrack0,n-1]$ labels the $n^{N}$
states and $\omega_n$ is given by eq. (\ref{Udef}). Orthonormality and
completeness read%

\begin{equation}
_{\mathbf{d}}\langle Cat\left|  Cat\right\rangle _{\mathbf{c}}=\delta
_{\mathbf{c},\mathbf{d}}\;,\qquad\sum_{\mathbf{c}}\left|  Cat\right\rangle
_{\mathbf{c}\,\,\mathbf{c}}\langle Cat\,|=\hat{1}\;.
\end{equation}

In the
low dimensional case of $N=n=2$ the definition (\ref{catdef}) reduces to the bell basis and
for $N=3$, $n=2$ the GHZ state $\frac{1}{\sqrt{2}}(\left|  000\right\rangle
+\left|  111\right\rangle )$ is one member of such a set.
Further examples for cat-states are given in Appendix C.

The characterization of these generalized cat states in terms of
cluster sums shows (cf. Table \ref{csum})

\begin{align}
Y_{1} &  =0\;,\\
Y_{m} &  =\frac{(n-1)^{m}+(-1)^{m}(n-1)}{n}\;,\qquad1\leq m<N\;,\\
Y_{N} &  =(n-1)\,n^{N-1}+\frac{(n-1)^{N}+(-1)^{N}(n-1)}{n}\;,
\label{maxYN}
\end{align}

while the purity factor distribution for any such a cat state is
(cf. eq. (\ref{PurityFactor}))

\begin{align}
p_{1}  &  =0\;,\qquad\qquad\text{(totaly mixed)}\\
p_{m}  &  =\frac{n^{m-1}-1}{n^{m}-1}\;,\qquad1<m<N\\
p_{N}  &  =1\;,\qquad\qquad\text{(pure)}\;.
\end{align}

\begin{figure}
\centering
\psfrag{n}{$n$} 
\psfrag{m}{$m$} 
\psfrag{pm}{$p_m$} 
\includegraphics[height=7cm]{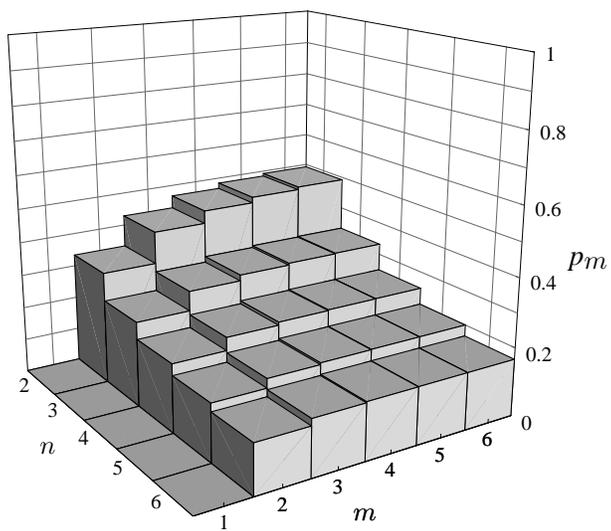}
\caption{$m$-cluster purity factor $p_{m}$, as defined in
eq. (\ref{PurityFactor}), for cat-states depending on the subsystem
dimension $n$ and the cluster size $m$ for $m<N$. The total cat-state
is a pure state, so $p_N$ always equals $1$.}
\label{PurityFactorDistribution}
\end{figure}

The purity factor distribution for cat-states is shown in
Fig. \ref{PurityFactorDistribution}.

For $n \gg 1$ any cluster, regardless of its size $m$, $(m<N)$
approaches the same purity factor $p_m=\frac{1}{n}$. This behaviour is
reminescent of that for the cluster-entropy.

For $N\rightarrow \infty$ the cluster-sum $Y_N$ almost exploits the sum
rule (\ref{sumrule}):
\begin{equation}
\frac{Y_N}{n^N} \rightarrow \frac{n-1}{n}
\end{equation}

Cat states thus become even more "non-classical" with increasing $N$
(and $n$).

\subsection{Collective invariants and eigenstates}

We have already briefly discussed invariants under general unitary
transformations ($\tr{\hat{\rho}^2}$) and under local unitary
transformations (the cluster sums). Let our
quantum network be described by the Hamiltonian $\hat{H}$, with the
spectral representation

\begin{equation}
\hat{H}=\sum_k E_k \hat{\rho}_k
\end{equation}

where $\hat{\rho}_k=\ket{\phi_k}\bra{\phi_k}$. Then the
expectation-values

\begin{equation}
C_j = \tr{\hat{\rho} \hat{\rho}_j} =
\bra{\phi_j}\hat{\rho}\ket{\phi_j}
\label{cj}
\end{equation}

constrained by $\sum_j C_j=1$ are additional invariants of motion under the unitary transformation
generated by $\hat{H}$. This follows from the fact, that $i \hbar
\frac{\partial}{\partial t} \tr{\hat{\rho} \hat{A}}=
\tr{{[\hat{H},\hat{\rho}]}_- \hat{A}} =\tr{{[\hat{A},\hat{H}]}_-
\hat{\rho}}$ which vanishes if $\hat{A}$ commutes with $\hat{H}$.
Note that $\ket{\phi_j}$ has, in general, not product form. We
consider 3 examples:

\begin{itemize}

\item[i.]
 $N=n=2$ network with F\"orster-interaction $C_{\mbox{\rm F}}$

The Hamiltonian in terms of the collective operators (cf. Sect. \ref{CollectiveOperators})

\begin{equation}
\hat{H}=\frac{\hbar\, \omega}{2}\, \hat{E}_{001,0}+\frac{\hbar\, C_{\mbox{\rm F}}}{2}\, (\hat{E}_{200,0}+\hat{E}_{020,0})
\end{equation}

has the eigenstates

\begin{eqnarray}
\ket{\phi_0}   &=& \ket{00} \label{es1}\\
\ket{\phi_{\pm}}&=& \frac{1}{\sqrt{2}}(\ket{01} \pm \ket{10})\\
\ket{\phi_1}   &=& \ket{11} \label{es3}
\end{eqnarray}

which can be rewritten as

\begin{eqnarray}
\hat{\rho}_0 & = &
\frac{1}{4}(\hat{1}+\hat{E}_{002,0}-\hat{E}_{001,0})\\
\hat{\rho}_{\pm} & = &
\frac{1}{4}(\hat{1}-\hat{E}_{002,0}\pm(\hat{E}_{200,0}+\hat{E}_{020,0}))\\
\hat{\rho}_1 & = &
\frac{1}{4}(\hat{1}+\hat{E}_{002,0}+\hat{E}_{001,0})
\end{eqnarray}

These are all permutation-symmetric density-operators; their expectation
values $C_j$ according to eq. (\ref{cj}) lead to three independent collective invariants

\begin{eqnarray}
E_{001,0} & = & \mbox{\rm const.}\\
E_{002,0} & = & \mbox{\rm const.}\\
E_{200,0}+E_{020,0} & = & \mbox{\rm const.}\
\end{eqnarray}

\item[ii.]
$N=n=2$ with renormalization interaction $C_{\mbox{\rm R}}$

The Hamiltonian
\begin{eqnarray}
\hat{H}&=&\frac{\hbar}{4}\,(\omega^{(1)}+\omega^{(2)}) \hat{E}_{001,0}\label{renorm}\\
 &&+
\frac{\hbar}{4}\,(\omega^{(1)}-\omega^{(2)}) \hat{E}_{001,1} +
\frac{\hbar\, C_{\mbox{\rm R}}}{2} \hat{E}_{002,0}\nonumber
\end{eqnarray}

has the product states as eigenstates. The collective invariants are
in this case

\begin{eqnarray}
E_{001,0} & = & \mbox{\rm const.}\\
E_{002,0} & = & \mbox{\rm const.}\\
E_{001,1} & = & \mbox{\rm const.}\
\end{eqnarray}

\item[iii.]
$N=n=2$ with collective stimulation

The Hamiltonian given by
\begin{equation}
\hat{H}=\frac{\hbar}{2}\, g\;\hat{E}_{100,0}+\frac{\hbar}{2}\, \delta\;\hat{E}_{001,0}
\label{specialH}
\end{equation}
describes a non-interacting spin-pair driven by a coherent
electromagnetic field (strength $g$, detuning $\delta$) in rotating
wave approximation. Its invariants
turn out to be

\begin{eqnarray}
\delta E_{001,0} + g E_{100,0}   =   \mbox{\rm const.}\\
\delta^2 E_{002,0}+ 2 g \delta E_{101,0} +  g^2 E_{200,0}   =  \mbox{\rm const.}\\
4 \delta^2 (g^2 + \delta^2) E_{001,1} - g^4 E_{002,0}\nonumber\\
-g^2(g^2 + \delta^2) E_{020,0}+ 4 g \delta (g^2+\delta^2)
 E_{100,1}\nonumber\\
+2 g^3 \delta E_{101,0}-g^2 \delta^2 E_{200,0}  =  \mbox{\rm const.}\label{var3}
\end{eqnarray}

\end{itemize}

Contrary to example ii, the $\hat{H}$-model (\ref{specialH}) is
permutation-symmetric, while the invariant (\ref{var3}) contains
expectation values with $b\neq0$. The latter can be unequal zero only
if at some previous stage of the preparation the permutation symmetry
had been broken (cf. Sect. \ref{interplay}).

\subsection{Permutation symmetry}

A complete description of large quantum networks turns out to be virtually
impossible, as the Hilbert space dimension grows exponentially with the number of
particles used. However, symmetry can reduce
the complexity of the total system by a significant amount: For a perfect
permutation-symmetric system we will show that the reduction in the number of
parameters needed is so enormous, that a polynomial increasing number of
parameters is enough to describe networks of arbitrary size.  

Within such a network no subsystem can be
distinguished, neither in preparation nor in detection. Two possible setups can be
thought of: Fundamental indistinguishable and operational
indistinguishable subsystems. A typical setup of the former type would be
several electrons in a box. As the electrons are fermions and their location
is not fixed, there is no way to act independently on a specific electron. All
controlling and measurement procedures act on the system as a whole. Therefore
no information loss can occur when reducing the system-description to
permutation-symmetric operators.

Opposed to that, in the case of operational indistinguishability, the design of the experimental setup is the source of reduction. A linear ion
trap with a laser beam acting on the ions could be used as an example. Only if
the beam waist is less than the spatial separation of the ions, the particles
become distinguishable, so by controlling the laser beam the experimentalist can
choose which case he is in.

Collective operators $\hat{E}_{\alpha\beta\gamma,b}$ with $b=0$
(cf. Sect. \ref{CollectiveOperators}) are invariant under any
arbitrary subsystem 
permutation $\hat{P}$%

\begin{equation}
\hat{P}^{\dagger}\hat{E}_{\alpha\beta\gamma,0} \hat{P} = \hat{E}_{\alpha\beta\gamma,0}\;.
\end{equation}
For indistinguishable subsystems, these are the only
operators allowed: Any control or measurement operator has to be
part of the Liouville-space spanned by
$\hat{E}_{\alpha\beta\gamma,0}$. The number of such operators is (for $n=2$)

\begin{equation}
\xi_0=\frac{(N+1)(N+2)(N+3)}{6}\;,
\label{xi}
\end{equation}

and thus scales polynomially of order
${\mathcal{O}}(N^3)$ with the number of subsystems involved. Therefore, such a
highly symmetric system constitutes a class of reduced complexity as
compared with a general quantum network of the same size $N$. This is
somewhat surprising as indistinguishability might have been expected
to enhance non-classical features. 

Alternatively, if all expectation-values happen to be permutation-symmetric (no
"structure"), then
\begin{equation}
\langle\hat{C}_{\alpha\beta\gamma,p}\rangle= \langle\hat{C}_{\alpha\beta\gamma}\rangle
\end{equation}
for all $p$-permutations, and by recalling eq. (\ref{omegasum}) the
definition eq. (\ref{defineE}) leads to
\begin{equation}
\langle\hat{E}_{\alpha\beta\gamma,b}\rangle=\Omega \langle\hat{C}_{\alpha\beta\gamma}\rangle \delta_{0b}
\end{equation}
i.e. we need to consider
only  the $\hat{E}_{\alpha\beta\gamma,0}$ operators. This would hold,
correspondingly, for the operators $\hat{F}_{z,\gamma,b}$ according to
eq. (\ref{Fdef}) (or $\hat{G}_{m,b}$ according to eq. (\ref{Gdef}))
provided only $z$, $\gamma$ (or $m$) could indeed be
distinguished. The number of remaining operators is then further
reduced: For $\hat{F}_{z,\gamma,0}$ this number is $\sum_{\gamma=0}^N
(2 (N-\gamma)+1)=(N+1)^2$, for $\hat{G}_{m,0}$ it is simply $N+1$.

\subsubsection{Symmetry classes}

Permutation symmetry defines a kind of operational
indistinguishability between the subsystems of a given network. This
symmetry alone, however, would allow for more symmetry classes than
realized in nature by the fundamentally indistinguishable particles,
Fermions and Bosons, respectively. The spin-statistics-relation going
back to Pauli \cite{25,23} might be relaxed, if the particles or
subsystems are localized in different areas of real space \cite{24}. The
location index would, in principle, render these subsystems
distinguishable; however, for the following we assume that the actual
operators describing the network and its coupling to the outside world
are still permutation-symmetric, so that the corresponding
super-selection rules apply, as will be discussed below. This would
imply, e.g. that electrons, localized in different semiconductor
quantum dots, could live in the state-subspace of Bose-symmetry (or of
any other "para-boson" \cite{23} symmetry class as well; cf. Table \ref{symtable}), provided we are able to prepare
such states from some standard initial state: Directed transient
symmetry breaking could do this job (cf. Sect. \ref{interplay}).

Any symmetry type can be characterized by a Young diagram and is
equivalent to a irreducible representation of the permutation group
$S_N$ spanned by the basis vectors. As an example, Table \ref{symtable} shows all
possible symmetries for a $N=4$ particle system. Since a definite
angular momentum quantum number $j$ is assigned to every Young diagram, there
are $2j+1$ states of equal symmetry type but different energy
(i.e. different configuration). If the
system is subjected to permutation-symmetric operators only,
the super-selection rules prohibit any transition between different Young
tables. In case the system under consideration is in a specific state with angular momentum
$j$, the state space reached by applying any collective operator $\hat{E}_{ijk,0}$  
is of dimension $2j+1$. Therefore $(2j+1)^2$ parameters (expectation
values) are needed to describe the system. The total number of
parameters, $\sum_{j=0}^{N/2}(2j+1)^2=\frac{1}{6}(N+1)(N+2)(N+3)$ is
exactly the number $\xi_0$ of collective
operators $\hat{E}_{ijk,0}$ (see eq. (\ref{xi})).

\subsubsection{Structure and Hamilton-models}
For fundamentally indistinguishable particles, the subsystem index
$\mu$ has no physical meaning. However, it may happen that a specific
property is not only a good quantum number but a unique constant of
motion for any subsystem. Localization in real space is a pertinent example;
the subsystem-index $\mu$ is then mapped onto a
spatial position-index $R_{\mu}$. By this position any subsystem
becomes distinguishable, in principle. The phases (entering the
collective operators $\hat{E}$) get a physical meaning ("wave length"). An "operational"
indistinguishability remains, if the Hamiltonian describing the
network does still contain collective-operators of the type
$\hat{E}_{\alpha\beta\gamma,0}$ only.

Typical Hamilton-models include
$m=1,2$ -particle operators. The structure tends to break
permutation-symmetry for the localized states as the coupling usually depends on
the distance $|R_{\mu}-R_{\nu}|$. This partial selectivity can be
described as a perturbation via collective operators 
$\hat{E}_{\alpha\beta\gamma,b}$ with $b\neq 0$. 

If all pertinent distances could be made equal, the breaking of the permutation
symmetry would go to zero. For $N>D+1$ (in
$D$ dimensions) however, the interaction distances cannot all
be the same. A partial remedy is the introduction of a "quantum bus":
In this case the nodes do not interact directly but only indirectly
via a common degree of freedom. This degree of freedom could be a
central spin, but typically is implemented as a collective mode (like
the phonon mode of a cold ion trap \cite{18}).

\subsection{Symmetry breaking and irreversibility}

Constrained operations will lead, quantum mechanically, to selection
rules accompanied by a tremendous reduction of the state space
available to the system dynamics starting from a given initial
state. This situation needs to be distinguished from lack of control
(measurement data) implying lack of information (entropy $S>0$). In so
far as this lack of control refers to expectation values $b\neq0$
(which would be absent for strict permutation symmetry), "uncontrolled" symmetry
breaking may be said to lead to an ensemble description and
irreversibility: The true state space has been reduced to the smaller
one defined by the assumed permutation symmetry. In this case,
however, the "decoherence" does not reflect the influence of an
external bath but is rather of "internal" origin. Phenomenologically
ony may try to model these effects, as usual, via some decoherence times;
however, it is not clear yet under what conditions such a procedure
would be appropriate. Clearly it should, at most, work for
sufficiently large networks and only to the extent that the reduction
is really substantial (cf. ref. \cite{20}). The decoherence time would
then, in turn, allow to assess symmetry breaking effects in a global
way.

In the following, we will restrict ourselves to unitary transformations.

\subsection{Special unitary transformations}
\subsubsection{Cyclic permutations: generalized echoes}
\label{cyclic}
The pulse-like manipulation of Hamiltonians,%

\begin{equation}
\hat{H}(t)=\hat{H}_{j}\;, \qquad\mbox{for } t_{j}\le t < t_{j+1}\;,
\quad j=0,1,2,\ldots
\end{equation}

to shape unitary evolutions%

\begin{equation}
\hat{U}_{j}(\Delta t_{j})=e^{\frac{-\hat{H}_{j}\Delta
t_{j}}{\hbar}}\;,\qquad \Delta t_j=t_{j+1}-t_j
\end{equation}

at will, has become a popular approach to quantum computation \cite{26}. One basic
operation is the ``halting operation'', $\hat{U}_{j}(\Delta t_{j})=\hat{1}$,
which would require $\hat{H}_{j}\equiv0$. This condition cannot be realized,
in general. The well-known spin-echoes \cite{30} are formally based on time reversal,
thus undoing a unitary evolution over a time period $\Delta t/2$ within the
next period $\Delta t/2$:%

\begin{equation}
\hat{U}_{j}(-\Delta t/2)=\hat{U}_{j}^{\dagger}(\Delta t/2)
\end{equation}

This time-reversal cannot be implemented either; however, its effect can be
simulated in any discrete Hilbert-space of dimension $n$ by means of the
cyclic permutation operations $\hat{U}_{n-1,0}$, as introduced in section \ref{UnitaryOperators}.

Let us consider the time-independent Hamiltonian $\hat{H}$. It is
convenient to require $\operatorname*{tr}\left\{  \hat{H} \right\} =
0$ so that

\begin{equation}
\sum_{k}^{n-1} E_{k}=0 \label{energiesum}
\end{equation}

with $E_{k}$ denoting the eigenvalues of $\hat{H}$. Then the effect of the
unitary time evolution generated by $\hat{H}$

\begin{equation}
\left|  \psi(\Delta t)\right\rangle = \hat{U}_{\hat{H}}(\Delta t) \left|
\psi(0)\right\rangle
\end{equation}

can be suppressed for any given $\Delta t$ by the iteration%

\begin{equation}
\hat{U}_{\mbox{\small eff}}(\Delta t):={\left(  \hat{U}_{n-1,0}\hat{U}_{\hat{H}}%
(\Delta t/n)\right)  }^{n} \label{mapping}
\end{equation}

Here we have assumed for simplicity that the application of $\hat{U}_{n-1,0}$
(cf. eq. (\ref{Udef})) does not consume any additional time. $\hat{U}_{n-1,0}$ generates cyclic
permutations between all the states $E_{k}$ of the spectrum. Any initial
eigenstate thus visits all the other eigenstates for the same time $\Delta
t/n$ so that, due to the constraint eq. (\ref{energiesum}) the total acquired phase adds up to
zero. This invariance property holds for any initial state $\left|
\psi(0)\right\rangle $, as such a state can be written as a superposition of
eigenstates, and any eigenstate returns to its initial phase separately.  

This mapping, eq. (\ref{mapping}), can be repeated to get the
stroboscopic invariance (cf. ref. \cite{11})

\begin{equation}
\hat{U}_{\mbox{\small eff}}(m\Delta t)={\left(  \hat{U}_{n-1,0}\hat{U}_{0}%
(\Delta t/n)\right)  }^{nm}=\hat{1}\;,
\end{equation}

where any initial state is periodically recovered. The cyclic permutation
$\hat{U}_{n-1,0}$ can be decomposed into a sequence of $n-1$ state-selective $\pi$-pulses;
for $n=3$, e.g.:%

\begin{eqnarray}
\hat{U}_{2,0}&=&\hat{P}_{01}+\hat{P}_{12}+\hat{P}_{20}\\
&=&(\hat{P}_{00}+\hat
{P}_{12}+\hat{P}_{21})\cdot(\hat{P}_{01}+\hat{P}_{10}+\hat{P}_{22})\nonumber
\end{eqnarray}

The total number of such $\pi$-pulses for $\U{eff}(m \Delta t)$ is $n(n-1)m$. The conventional spin-echo
obtains for $n=2$, $m=1$, and an initial state $\left|  \psi(0)\right\rangle $
generated from the ground state by means of a $\pi/2$-pulse. The second
$\hat{U}_{1,0}$-flip is then usually omitted. 

\subsubsection{Selective control}

In principle, the results obtained in Sect. \ref{cyclic} may also be used for a network of
$N$ subsystems of dimension $n$ each. The condition is, that one is able to
implement the respective cyclic operator: For a network with
(distinguishable)
non-interacting subsystems we simply have to replace $\hat{U}_{n-1,0}$ by the
product-operator $\hat{Q}=\hat{U}_{n-1,0}^{(1)}\otimes\ldots\otimes\hat{U}_{n-1,0}^{(N)}$. 

In the case of identical subsystems, interactions may lift the
degeneracy of the network-eigenstates. We consider the example of a
network of $2$-level-subsystems and an
interaction of the type%

\begin{equation}
\hat{H}^{\prime}=\hbar\,\sum_{\mu<\nu}C_{R}^{\mu\nu} \hat{\lambda}_{3}^{(\mu)}%
\otimes\hat{\lambda}_{3}^{(\nu)}\;.
\end{equation}

This interaction modifies the eigen-spectrum (allowing for
selectivity) while leaving the eigenstates
unchanged (i.e. as product-states), $\left|
p(N)\ldots p(1)\right\rangle $; $p(\mu)=0,1$ (cf. eq. (\ref{renorm})). Single-particle transitions are
allowed only between states which differ at one position (``Hamming distance''
1). A convenient numbering, $s$, of the $2^{N}$ states allowing for cyclic
permutation is given in Table \ref{table2}. For a large network $N$, the echo becomes
difficult to implement practically, as it requires $\sim 2^{2N}$ short but
at the same time frequency-selective $\pi$-pulses per time period
$\Delta t$. In general, interactions will also modify the eigenstates (i.e. lead
to a non-product form). A simple example for $N=n=2$ is provided by
eq. (\ref{es1}) - (\ref{es3}).

It has been proposed \cite{2} to exploit some
approximate stroboscopic invariance also for open systems under the
condition  $\Delta t \ll \tau_c$, where $\tau_c$ is a typical
time-scale for dissipation. However, such short pulses will easily come
in conflict with their required selectivity in frequency-space and
thus tend to restrict practical implementations of their proposal to
small networks only. Furthermore, it is unlikely that the additional
control interactions could work without inducing uncontrollable
features on their own. Supression of decoherence would amount to
increase the pertinent decoherence-time and -length to macroscopic
dimensions (which typically happens for very special and
restricted state-spaces only, like in superconductivity or Bose-Einstein
condensation).

\begin{table}
\begin{tabular}
[c]{c|rrrrrrrrrl}%
$N\backslash s$ & 0 & 1 & 2 & 3 & 4 & 5 & 6 & 7 & 8 & \ldots\\\hline
1 & 0 & 1 &  &  &  &  &  &  &  & \\
2 & 00 & 01 & 11 & 10 &  &  &  &  &  & \\
3 & 000 & 001 & 011 & 010 & 110 & 111 & 101 & 100 &  & \\
4 & 0000 & 0001 & 0011 & 0010 & 0110 & 0111 & 0101 & 0100 & 1100 & \ldots
\end{tabular}
\caption{Circular sequences of product states with Hamming distance 1
for various network sizes $N$. These sequences can be
constructed recursively: Sequence $N+1$ is obtained from that of $N$ by
repeating each member and supplementing it from the right by $0,1$ $(1,0)$,
respectively. (There are other sequences with the same property.)}
\label{table2}
\end{table}

\subsubsection{Collective control}
We consider the unitary evolution induced by a "collective" and
permutation-symmmetric Hamiltonian like

\begin{equation}
\hat{H}  =  \hbar\, \alpha\, \hat{E}_{m00,0}=\hbar\, \alpha \sum_{\mu<\nu\ldots<\xi}
\hat{\sigma}_x^{(\mu)} \hat{\sigma}_x^{(\nu)}\ldots
\hat{\sigma}_x^{(\xi)}
\end{equation}

This $\hat{H}$ generates the unitary time evolution

\begin{eqnarray}
\hat{U}(t)  &=&  e^{-i \alpha t \hat{E}_{m00,0}}\label{Ut}\\
&=&\prod_{\mu<\nu\ldots<\xi}(\cos(\alpha t) \, \hat{1} - i\, \sin(\alpha
t)\,
\hat{\sigma}_x^{(\mu)}\hat{\sigma}_x^{(\nu)}\ldots\hat{\sigma}_x^{(\xi)}
)\;.
\nonumber
\end{eqnarray}

In particular for $m=1$ and $\alpha t=\frac{\pi}{2}$ on finds
$\hat{U}_{\frac{\pi}{2}}={(-i)}^N \hat{E}_{N00,0}$, while for $\alpha
t=\frac{\pi}{4}$, $\hat{U}_{\frac{\pi}{4}}\propto\sum_{\mu=0}^N {(-i)}^{\mu} \hat{E}_{\mu00,0}$.

For $m=2$ and $\alpha t=\frac{\pi}{2}$, we have
$\hat{U}_{\frac{\pi}{2}} = (-i)^{\frac{N}{2}} \hat{E}_{N00,0}$ for $N$
even and $\hat{U}_{\frac{\pi}{2}}=i^{\frac{N-1}{2}} \hat{1}$ otherwise; for $N$ even
and time $\alpha t=\frac{\pi}{4}$, eq. (\ref{Ut}) simplifies to 

\begin{align}
\hat{U}_{\frac{\pi}{4}} \propto \id + i \hat{E}_{N00,0}\;, \quad
&\mbox{\rm for } \frac{N}{2}\;\mbox{\rm  even}\\ 
\hat{U}_{\frac{\pi}{4}} \propto \id - i \hat{E}_{N00,0}\;, \quad
&\mbox{\rm for } \frac{N}{2}\;\mbox{\rm  odd} 
\end{align}




 Applied to the ground state
$\ket{00\ldots0}$, this $\hat{U}_{\frac{\pi}{4}}$ thus creates the
$N$-particle cat state (cf. Sect. \ref{generalizedCats} and Appendix C)

\begin{equation}
\hat{U}_{\frac{\pi}{4}} \ket{00\ldots 0}  \propto \ket{00\ldots 0}
\pm i \ket{11\ldots 1}\;
\end{equation}

in a single step! Specific examples for such collective control
scenarios have been discussed in Refs. \cite{1,7,8,9,10,17}.

\subsubsection{Interplay between selective and collective
interactions}
\label{interplay}

Selective interactions violate the selection rules implicit in
permutation-symmetric interactions. This qualitatively different
dynamical behaviour can be exploited to implement specific
functionalities: One possibility is to address different
symmetry-classes via selective coupling (i.e. controlled symmetry breaking); another possibility is to
suppress transition due to permutation-symmetric interactions by using
states of different symmetry, which, nevertheless, could all be
prepared selectively ("decoherence-free subspace").

To be specific, let us consider a network of $N$ pseudo-spins without
mutual interactions, but in the presence of a quantum bus (i.e. a
collective mode to which all spins are coupled in the same way). We
further assume that the coupling to a larger field can be made at will
either selective or collective. We start (for $N=4$, $n=2$) with the
permutation-symmetric ground state $\ket{0000}$, ($j=2$). By
applying selective laser pulses and exploiting the coupling to the
quantum-bus we can generate the EPR-state
$\frac{1}{\sqrt{2}}\,(\ket{01}-\ket{10})$ within the pair
$\nu=1,2$. The total 4-particle state then becomes a member of the
symmetry class $j=1$, see Table \ref{symtable}. If permutation
symmetry is restored now, transitions are possible only within the
$(2j+1)$-dimensional subspace of this symmetry class (on a time-scale
less than the decoherence time). This would allow to study finite
systems of (operationally) indistinguishable subsystems of a symmetry
not realized in nature by fundamentally indistinguishable particles! It is straightforward to extend this scheme to
$N>4$.

Alternatively, the selection rules can readily be exploited as a means
for stabilization: For this purpose we assume that the "unwanted"
dynamics (coupling to the bath) is permutation-symmetric, while the
control dynamics to be used is selective (i.e. not subject to the
selection rules). For $N=4$ we could take the three lowest energy states
$j=2,1,0$. Unitary dynamics within this subspace would (in the
ideal case) not be perturbed by dissipation, if the bath was kept at
zero temperature. (There are higher energy levels with the same
symmetry, though.)

It is thus preferable to use the multiplicity of the states $j=0$ (for
$N$ even) which would decouple from the bath exactly, as each of those
states is the only member of its symmetry class (see Table \ref{symtable}): There is nothing to
connect to under the action of a permutation-symmetric coupling
(cf. ref \cite{14}). Such schemes have been investigated by a
number of authors \cite{5,6,12,13,15}. The resulting stabilization is
limited by the fact that the symmetry selection rules will, in
practice, not hold strictly.

\section{Summary and Conclusions}

In this paper we have addressed composite systems consisting of $N$
subsystems $\mu$ of respective dimension $n_{\mu}$. We have been concerned
with applications of (selective) cluster-operators and collective
operators, both based on unitary local operator sets, and both being
orthogonal and complete. The former are very useful in characterizing
entanglement of pure network states, in particular of cat-states, for
any $n$ and $N$.  Commuting sets of cluster-operators have been
derived. The latter are adapted to situations in which
individual subsystems cannot be selected. Classically this would imply
a loss of information about the resulting state. Quantum-mechanically, this
lack of control would rather
give rise to entanglement --  with super-selection rules
tremendously reducing the state-space available: Actually its dimension becomes
polynomial in $N$. Such states have been proposed as subspaces of
reduced decoherence (with respect to permutation-symmetric bath interations). We have suggested that
localized, artificial quantum networks, subjected to certain
symmetry-breaking initialization steps, might live in
state-symmetry-classes that otherwise cannot occur for fundamentally
indistinguishable particles.

\acknowledgments

We thank the Deutsche Forschungsgemeinschaft (Schwerpunkt
Quanteninformationsverarbeitung) for financial support.
\newpage

\appendix


\widetext
\onecolumn
\section{Unitary operator basis for $n=3$}

\[
\begin{array}[c]{ccc}
\rule[-0.8cm]{0cm}{1cm}\hat{U}_{00}=\left(
\begin{array}
[c]{ccc}%
1 & 0 & 0\\
0 & 1 & 0\\
0 & 0 & 1
\end{array}
\right)  & \hat{U}_{01}=\left(
\begin{array}
[c]{ccc}%
1 & 0 & 0\\
0 & e^{\frac{2\pi i}{3}} & 0\\
0 & 0 & e^{-\frac{2\pi i}{3}}%
\end{array}
\right)  & \hat{U}_{02}=\left(
\begin{array}
[c]{ccc}%
1 & 0 & 0\\
0 & e^{-\frac{2\pi i}{3}} & 0\\
0 & 0 & e^{\frac{2\pi i}{3}}%
\end{array}
\right) \\
\rule[-0.8cm]{0cm}{1cm}\hat{U}_{10}=\left(
\begin{array}
[c]{ccc}%
0 & 0 & 1\\
1 & 0 & 0\\
0 & 1 & 0
\end{array}
\right)  & \hat{U}_{11}=\left(
\begin{array}
[c]{ccc}%
0 & 0 & e^{-\frac{2\pi i}{3}}\\
1 & 0 & 0\\
0 & e^{\frac{2\pi i}{3}} & 0
\end{array}
\right)  & \hat{U}_{12}=\left(
\begin{array}
[c]{ccc}%
0 & 0 & e^{\frac{2\pi i}{3}}\\
1 & 0 & 0\\
0 & e^{-\frac{2\pi i}{3}} & 0
\end{array}
\right) \\
\hat{U}_{20}=\left(
\begin{array}
[c]{ccc}%
0 & 1 & 0\\
0 & 0 & 1\\
1 & 0 & 0
\end{array}
\right)  & \hat{U}_{21}=\left(
\begin{array}
[c]{ccc}%
0 & e^{\frac{2\pi i}{3}} & 0\\
0 & 0 & e^{-\frac{2\pi i}{3}}\\
1 & 0 & 0
\end{array}
\right)  & \hat{U}_{22}=\left(
\begin{array}
[c]{ccc}%
0 & e^{-\frac{2\pi i}{3}} & 0\\
0 & 0 & e^{\frac{2\pi i}{3}}\\
1 & 0 & 0
\end{array}
\right)
\end{array}
\]\newline

\narrowtext

\widetext
\section{Relations between operator sets}

The relations between this unitary operator set $\left\{  \hat{U}%
_{ab}\right\}  $, the SU($n$) generators $\left\{  \hat{u}_{jk},\hat{v}%
_{jk},\hat{w}_{l}\right\}  $ and the projection operators$\left\{  \hat
{P}_{ij}\right\}  $ are given by:

\begin{equation}
\hat{P}_{jk}=\frac{1}{n}\sum_{b=0}^{n-1}\omega_n^{-b k}\hat{U}_{\underline
{j-k},b}%
\end{equation}%

\begin{equation}
\hat{U}_{ab}=\sum_{k=0}^{n-1}\omega_n^{kb}\hat{P}_{\underline{a+k},k}%
\end{equation}%

\begin{align}
\hat{U}_{ab}  &  =\sum_{0\leq j<k< n}\frac{1}{2}(\omega_n^{b j}%
\delta_{\underline{j+a},k}+\omega_n^{b k}\delta_{\underline{k+a}%
,j})\,\hat{u}_{jk}+\\
&  \sum_{0\leq j<k< n}\frac{i}{2}(\omega_n^{b j}\delta_{\underline
{j+a},k}-\omega_n^{b k}\delta_{\underline{k+a},j})\,\hat{v}%
_{jk}+\nonumber\\
&  \sum_{0\leq l< n-1}-\frac{1}{\sqrt{2(l+1)(l+2)}}\,\delta_{0,a}%
\,(-(l+1)\,\omega_n^{b (l+1)}+\sum_{q=1}^{l+1}\omega_n^{b(q-1)})\,\hat{w}_{l}\nonumber
\end{align}%

\begin{align}
\hat{u}_{jk}  &  =\frac{1}{n}\sum_{a,b=0}^{n-1}(\omega_n^{-b k}%
\,\delta_{a,\underline{j-k}}+\omega_n^{-b j}\,\delta_{a,k-j})\,\hat{U}_{ab}\\
\hat{v}_{jk}  &  =\frac{i}{n}\sum_{a,b=0}^{n-1}(\omega_n^{-b k}%
\,\delta_{a,\underline{j-k}}-\omega_n^{-b j}\,\delta_{a,k-j})\,\hat{U}%
_{ab}\nonumber\\
\hat{w}_{l}  &  =-\frac{1}{n}\sqrt{\frac{2}{(l+1)(l+2)}}\sum_{b=0}^{n-1}%
(-(l+1)\,\omega_n^{-b (l+1)}+\sum_{q=1}^{l+1}\omega_n^{-b(q-1)})\,\hat{U}_{0,b}\nonumber
\end{align}
\newline
\narrowtext

\twocolumn
\section{Examples for generalized cat states}

We give some (low dimensional) examples for generalized cat states consisting of $N$
subsystems with each one being a $n$ level system. Together they form
a orthonormal and complete basis of the underlying $n^N$ dimensional Hilbert
space.\newline

$N=2$:

\bigskip\qquad$n=2$: \begin{list}{}{\parsep0.5ex}
\item$\qquad\qquad\left|  Cat\right\rangle_{00}=\frac{1}{\sqrt{2}}
\left(  \left|
00\right\rangle+\left|  11\right\rangle\right)  $
\item$\qquad\qquad\left|  Cat\right\rangle_{01}=\frac{1}{\sqrt{2}}
\left(  \left|
01\right\rangle+\left|  10\right\rangle\right)  $
\item$\qquad\qquad\left|  Cat\right\rangle_{10}=\frac{1}{\sqrt{2}}
\left(  \left|
00\right\rangle-\left|  11\right\rangle\right)  $
\item$\qquad\qquad\left|  Cat\right\rangle_{11}=\frac{1}{\sqrt{2}}
\left(  \left|
01\right\rangle-\left|  10\right\rangle\right)  $
\end{list}

\qquad$n=3$: \begin{list}{}{\parsep0.5ex}
\item$\qquad\qquad\left|  Cat\right\rangle_{00}=\frac{1}{\sqrt{3}}
\left(  \left|
00\right\rangle+\left|  11\right\rangle+\left|  22\right\rangle\right)  $
\item$\qquad\qquad\left|  Cat\right\rangle_{01}=\frac{1}{\sqrt{3}}
\left(  \left|
01\right\rangle+\left|  12\right\rangle+\left|  20\right\rangle\right)  $
\item$\qquad\qquad\left|  Cat\right\rangle_{02}=\frac{1}{\sqrt{3}}
\left(  \left|
02\right\rangle+\left|  10\right\rangle+\left|  21\right\rangle\right)  $
\item$\qquad\qquad\left|  Cat\right\rangle_{10}=\frac{1}{\sqrt{3}}
\left(  \left|
00\right\rangle+e^{\frac{2}{3}\pi i}\left|  11\right\rangle+e^{-\frac{2}{3}\pi
i}\left|  22\right\rangle\right)  $
\item$\qquad\qquad\left|  Cat\right\rangle_{11}=\frac{1}{\sqrt{3}}
\left(  \left|
01\right\rangle+e^{\frac{2}{3}\pi i}\left|  12\right\rangle+e^{-\frac{2}{3}\pi
i}\left|  20\right\rangle\right)  $
\item$\qquad\qquad\left|  Cat\right\rangle_{12}=\frac{1}{\sqrt{3}}
\left(  \left|
02\right\rangle+e^{\frac{2}{3}\pi i}\left|  10\right\rangle+e^{-\frac{2}{3}\pi
i}\left|  21\right\rangle\right)  $
\item$\qquad\qquad\left|  Cat\right\rangle_{20}=\frac{1}{\sqrt{3}}
\left(  \left|
00\right\rangle+e^{-\frac{2}{3}\pi i}\left|  11\right\rangle+e^{\frac{2}{3}\pi
i}\left|  22\right\rangle\right)  $
\item$\qquad\qquad\left|  Cat\right\rangle_{21}=\frac{1}{\sqrt{3}}
\left(  \left|
01\right\rangle+e^{-\frac{2}{3}\pi i}\left|  12\right\rangle+e^{\frac{2}{3}\pi
i}\left|  20\right\rangle\right)  $
\item$\qquad\qquad\left|  Cat\right\rangle_{22}=\frac{1}{\sqrt{3}}
\left(  \left|
02\right\rangle+e^{-\frac{2}{3}\pi i}\left|  10\right\rangle+e^{\frac{2}{3}\pi
i}\left|  21\right\rangle\right)  $
\end{list}

$N=3$:

\bigskip\qquad$n=2$: \begin{list}{}{\parsep0.5ex}
\item$\qquad\qquad\left|  Cat\right\rangle_{000}=\frac{1}{\sqrt{2}}
\left(  \left|
000\right\rangle+\left|  111\right\rangle\right)  $
\item$\qquad\qquad\left|  Cat\right\rangle_{001}=\frac{1}{\sqrt{2}}
\left(  \left|
001\right\rangle+\left|  110\right\rangle\right)  $
\item$\qquad\qquad\left|  Cat\right\rangle_{010}=\frac{1}{\sqrt{2}}
\left(  \left|
010\right\rangle+\left|  101\right\rangle\right)  $
\item$\qquad\qquad\left|  Cat\right\rangle_{011}=\frac{1}{\sqrt{2}}
\left(  \left|
011\right\rangle+\left|  100\right\rangle\right)  $
\item$\qquad\qquad\left|  Cat\right\rangle_{100}=\frac{1}{\sqrt{2}}
\left(  \left|
000\right\rangle-\left|  111\right\rangle\right)  $
\item$\qquad\qquad\left|  Cat\right\rangle_{101}=\frac{1}{\sqrt{2}}
\left(  \left|
001\right\rangle-\left|  110\right\rangle\right)  $
\item$\qquad\qquad\left|  Cat\right\rangle_{110}=\frac{1}{\sqrt{2}}
\left(  \left|
010\right\rangle-\left|  101\right\rangle\right)  $
\item$\qquad\qquad\left|  Cat\right\rangle_{111}=\frac{1}{\sqrt{2}}
\left(  \left|
011\right\rangle-\left|  100\right\rangle\right)  $
\end{list}

\medskip

The cat-states $N=n=2$ can entirely be expressed in terms of
2-particle collective operators:

\begin{eqnarray}
\hat{\rho}_{00} &=& \frac{1}{4}(\hat{1} + \hat{E}_{200,0} -
\hat{E}_{020,0} + \hat{E}_{002,0})\\
\hat{\rho}_{01} &=& \frac{1}{4}(\hat{1} + \hat{E}_{200,0} +
\hat{E}_{020,0} - \hat{E}_{002,0})\\
\hat{\rho}_{10} &=& \frac{1}{4}(\hat{1} - \hat{E}_{200,0} +
\hat{E}_{020,0} + \hat{E}_{002,0})\\
\hat{\rho}_{11} &=& \frac{1}{4}(\hat{1} - \hat{E}_{200,0} -
\hat{E}_{020,0} - \hat{E}_{002,0})
\end{eqnarray}

These are all permutation-symmetric and involve $N=2$-particle
operators only.
For $N>n=2$ collective operators appear also of order $<N$: For $N=3$,
$\ket{\mbox{\rm Cat}}_{000}=\frac{1}{\sqrt{2}}(\ket{000}+\ket{111})$,
e.g., one finds

\begin{equation}
\hat{\rho}_{000} = \frac{1}{2^3}(\hat{1} + \hat{E}_{300,0} -
\hat{E}_{120,0} + \hat{E}_{002,0})
\end{equation}

The last term is responsible for the surviving $2$-particle properties. In general,
the cat-states are no longer permutation-symmetric: A pertinent
example is

\begin{align}
\hat{\rho}_{001}  = &\frac{1}{2^3}(\hat{1} - \frac{1}{3} \hat{E}_{002,0} +
\frac{2}{3} \hat{E}_{002,1}+ \frac{2}{3} \hat{E}_{002,2}\\
 &+\frac{1}{3} \hat{E}_{120,0} + \frac{2}{3} e^{-i\frac{\pi}{3}}
\hat{E}_{120,1}+ \frac{2}{3} e^{-i\frac{\pi}{3}} \hat{E}_{120,2})\;.\nonumber
\end{align}


\newpage
\onecolumn
\newpage


\onecolumn

\widetext
\begin{table}
\begin{tabular}{|l|l|l|l|l|}
\hline
Conf.&$m$&$j$&Young-T.&Basis vector\\
\hline\hline
1111&2&2& \tabe 1234  & $\ket{1111}$\\ \hline
0111&1&2& \tabe 1234  & $\frac{1}{2} (\ket{1110}+\ket{1101}+\ket{1011}+\ket{0111})$\\ \cline{3-5}
&&1& \tabf 1234  & $\frac{1}{2 \sqrt 3} (3
\ket{1110}-\ket{1101}-\ket{1011}-\ket{0111})$\\ \cline{4-5}
&& & \tabf 1243  & $\frac{1}{\sqrt 6} \left(2
   \ket{1101}-\ket{1011}-\ket{0111}\right)$\\ \cline{4-5}
&& & \tabf 1342  & $\frac{1}{\sqrt 2} \left(\ket{0111}-\ket{1011}\right)$\\ \hline
0011&0&2 & \tabe 1234 & $\frac{1}{\sqrt 6}
(\ket{0011}+\ket{0101}+\ket{0110}+\ket{1001}+\ket{1010}+\ket{1100})$\\ \cline{3-5}
&&1& \tabf 1234 & $\frac{1}{\sqrt 6} (\ket{0011}+\ket{0101}-\ket{0110}+\ket{1001}-\ket{1010}-\ket{1100})$\\ \cline{4-5}
&& & \tabf 1243 & $\frac{1}{2 \sqrt 3} (2
 \ket{0011}-\ket{0101}+\ket{0110}-\ket{1001}+\ket{1010}-2 \ket{1100})$\\ \cline{4-5}
&& & \tabf 1342  & $\frac{1}{2}
 (\ket{0101}+\ket{0110}-\ket{1001}-\ket{1010})$\\ \cline{3-5}
&&0 & \tabg 1234 & $\frac{1}{2}
 (\ket{0011}-\ket{0110}-\ket{1001}+\ket{1100})$\\ \cline{4-5}
&& & \tabg 1324  & $\frac{1}{2}
 (\ket{0101}-\ket{0110}-\ket{1001}+\ket{1010})$\\ \hline
0001&-1&2 & \tabe 1234  & $\frac{1}{2} (\ket{0001}+\ket{0010}+\ket{0100}+\ket{1000})$\\ \cline{3-5}
&&1 & \tabf 1234  & $\frac{1}{2 \sqrt 3} (3
\ket{0001}-\ket{0010}-\ket{0100}-\ket{1000})$\\ \cline{4-5}
&& & \tabf 1243  & $\frac{1}{\sqrt 6} \left(2
   \ket{0010}-\ket{0100}-\ket{1000}\right)$\\ \cline{4-5}
&& & \tabf 1342  & $\frac{1}{\sqrt 2}
\left(\ket{1000}-\ket{0100}\right)$\\ \hline
0000&-2&2 & \tabe 1234  & $\ket{0000}$\\ \hline
\end{tabular}
\caption{Symmetry classes for $N=4$ subsystems. The configuration
reflects the number of spins up or down and therefore the energy level
$m$. The angular momentum quantum number $j$ defines the symmetry
class with respect to permutation. The basis vectors are uniquely labeled by their
Young tableau together with the magnetic quantum number $m$. For
permutation-symmetric operators, super-selection rules only
allow transitions between states with the same Young tableau.}
\label{symtable}
\end{table}
\narrowtext


\begin{references}
\bib{27}{A. Bohr and O. Ulfbeck}{}{Rev. Mod. Phys. \vol{67}, 1 (1995).}
\bib{21}{G. Mahler and V. A. Weberrus}{Quantum Networks: Dynamics of
Open Nanostructures}{Springer New York, Berlin, 2nd edition (1998).}
\bib{26}{A. Steane}{}{Repts. Progr. Phys. \vol{61}, 117 (1998).}
\bib{22}{G. Mahler, M. Keller and R. Wawer}{}{Z. Physik, B104, 153 (1997).}

\bib{16}{E. Knill}{Non-Binary Unitary Error Bases and Quantum
Codes}{quant-ph/9608048 (1996).}
\bib{16z}{E. Knill and R. Laflamme}{}{Phys. Rev. A \vol{55}, 900 (1997).}
\bib{28}{J. Schlienz and G. Mahler}{Description of Entanglement}{Phys. Rev. A \vol{52}, 4396 (1995).}
\bib{29}{V. Vedral, M. B. Plenio, M. A. Rippin and P. L. Knight}{}{Phys. Rev. Lett.
\vol{87}, 2275 (1997).}
\bib{19}{M. Munowitz and M. Mehring}{}{Solid State Commun. \vol{64},
605 (1987).}
\bib{31}{D. I. Fivel}{}{Phys. Rev. Lett. \vol{74}, 835 (1995).}
\bib{25}{{\rm See, e.g. }R. F. Streater, A. S. Wightman}{PCT, Spin and
Statistics and All That}{W A. Benjamin, N.Y. (1964).}
\bib{24}{A. S. Alexandrov and R. T. Giles}{Parastatistics of charged
bosons partly localized by impurities}{cond-mat/9704164 (1997).}
\bib{23}{O. W. Greenberg and A. M. L. Messiah}{Selection Rules for
Parafields and the absence of para particles in Nature}{Phys. Rev. 138B, 1155
(1965).}
\bib{18}{J. I. Cirac and P. Zoller}{QC with cold trapped
ions}{Phys. Rev. Lett. \vol{74}, 4091 (1995).}
\bib{20}{M. Munowitz, A. Pines and M. Mehring}{}{J. Chem. Phys. \vol{86}, 3172 (1987).}
\bib{11}{G. Havel and V. M. Akulin}{Complete control of Hamiltonan Quantum systems: Engineering of Floquet evolution}{Phys. Rev. Lett. \vol{82}, 1
(1999).}
\bib{30}{R. Kimmrich}{NMR: Tomography, Diffusometry,
Relaxometry}{Springer Berlin, New York (1997).}
\bib{2}{L. Viola, E. Knill and S. Lloyd}{Dynamical decoupling of open
quantum systems}{ Phys. Rev. Lett. \vol{82}, 2417 (1999).}
\bib{1}{E. S. Polzik}{EPR-correlated atomic ensembles}{Phys. Rev. A \vol{59},
4202 (1999).}
\bib{7}{K. M{\o}lmer and A. S{\o}rensen}{Multi-particle entanglement of hot trapped ions}{Phys. Rev. Lett. \vol{82}, 1835 (1999).}
\bib{8}{A. S\o rensen and K. M\o lmer}{Quantum Computation with Ions in thermal motion}{Phys. Rev. Lett. \vol{82}, 1971 (1999).}
\bib{9}{A. S\o rensen and K. M\o lmer}{Spin-spin interaction and spin-squeezing in an optical lattice}{quant-ph/9903044 (1999).}
\bib{10}{D. Jaksch, M. J. Briegel, J. I. Cirac, C. W. Gardiner and
P. Zoller}{Entanglement of atoms via cold controlled collisions}{Phys. Rev. Lett. \vol{82}, 1975 (1999).}
\bib{17}{L. Quiroga and N. F. Johnson}{Entangled Bell and GHZ states
of excitons in coupled quantum dots}{cond-mat/9901201 (1999).}
\bib{14}{P. Zanardi and M. Rasetti}{Noiseless Quantum Codes}{Phys. Rev. Lett.
\vol{79}, 3306 (1997).}
\bib{5}{D. A. Lidar, I. L. Chuang and K. B. Whaley}{Decoherence-free
subspaces for QC}{Phys. Rev. Lett. \vol{81}, 2594 (1998).}
\bib{6}{P. Zanardi and F. Rossi}{Subdecoherent information encoding in a quantum dot array}{Phys. Rev. B \vol{59}, 8170 (1998).}
\bib{12}{L. Duan and G. Guo}{Preventing of dissipation with 2
particles}{Phys. Rev. A \vol{57}, 2399 (1998).}
\bib{13}{L. Duan and G. Guo}{Scheme for reducing collective
decoherence in quantum memory}{Phys. Lett. A \vol{243}, 265 (1998).}
\bib{15}{P. Zanardi}{Symmetrizing Evolutions}{quant-ph/9809064 (1998).}



\end{references}
\end{document}